\documentclass[3p]{elsarticle}
\usepackage{amssymb}
\usepackage{amsmath}
\usepackage{graphicx}
\usepackage{subfig}
\usepackage{psfrag}
\usepackage{pstool}
\usepackage{color}
\usepackage{subfig}
\usepackage[official]{eurosym}
\usepackage{textcomp}

\biboptions{comma,square}

\providecommand\matlabtextAA{\color[rgb]{0.000,0.000,0.000}\fontsize{8}{8}\selectfont\strut}%
\psfrag{0067}[tc][tc]{\matlabtextAA $\log_{10}v$}%
\psfrag{0068}[cl][cl]{\matlabtextAA 23-02-2007}%
\psfrag{0069}[tc][tc]{\matlabtextAA $\log_{10}v$}%
\psfrag{0070}[cl][cl]{\matlabtextAA 22-02-2007}%
\psfrag{0071}[cl][cl]{\matlabtextAA 21-02-2007}%
\psfrag{0072}[cl][cl]{\matlabtextAA 20-02-2007}%
\psfrag{0073}[cl][cl]{\matlabtextAA 19-02-2007}%
\def\matlabfragNegXTick{\mathord{\makebox[0pt][r]{$-$}}}
\providecommand\matlabtextAB{\color[rgb]{0.000,0.000,0.000}\fontsize{5}{5}\selectfont\strut}%
\psfrag{0000}[ct][ct]{\matlabtextAB $0$}%
\psfrag{0001}[ct][ct]{\matlabtextAB $1$}%
\psfrag{0002}[ct][ct]{\matlabtextAB $2$}%
\psfrag{0003}[ct][ct]{\matlabtextAB $3$}%
\psfrag{0004}[ct][ct]{\matlabtextAB $4$}%
\psfrag{0005}[ct][ct]{\matlabtextAB $5$}%
\psfrag{0006}[ct][ct]{\matlabtextAB $6$}%
\psfrag{0007}[ct][ct]{\matlabtextAB $7$}%
\psfrag{0008}[ct][ct]{\matlabtextAB $8$}%
\psfrag{0009}[ct][ct]{\matlabtextAB $9$}%
\psfrag{0010}[ct][ct]{\matlabtextAB $10$}%
\psfrag{0015}[ct][ct]{\matlabtextAB $0$}%
\psfrag{0016}[ct][ct]{\matlabtextAB $1$}%
\psfrag{0017}[ct][ct]{\matlabtextAB $2$}%
\psfrag{0018}[ct][ct]{\matlabtextAB $3$}%
\psfrag{0019}[ct][ct]{\matlabtextAB $4$}%
\psfrag{0020}[ct][ct]{\matlabtextAB $5$}%
\psfrag{0021}[ct][ct]{\matlabtextAB $6$}%
\psfrag{0022}[ct][ct]{\matlabtextAB $7$}%
\psfrag{0023}[ct][ct]{\matlabtextAB $8$}%
\psfrag{0024}[ct][ct]{\matlabtextAB $9$}%
\psfrag{0025}[ct][ct]{\matlabtextAB $10$}%
\psfrag{0026}[ct][ct]{\matlabtextAB $0$}%
\psfrag{0027}[ct][ct]{\matlabtextAB $1$}%
\psfrag{0028}[ct][ct]{\matlabtextAB $2$}%
\psfrag{0029}[ct][ct]{\matlabtextAB $3$}%
\psfrag{0030}[ct][ct]{\matlabtextAB $4$}%
\psfrag{0031}[ct][ct]{\matlabtextAB $5$}%
\psfrag{0032}[ct][ct]{\matlabtextAB $6$}%
\psfrag{0033}[ct][ct]{\matlabtextAB $7$}%
\psfrag{0034}[ct][ct]{\matlabtextAB $8$}%
\psfrag{0035}[ct][ct]{\matlabtextAB $9$}%
\psfrag{0036}[ct][ct]{\matlabtextAB $10$}%
\psfrag{0041}[ct][ct]{\matlabtextAB $0$}%
\psfrag{0042}[ct][ct]{\matlabtextAB $1$}%
\psfrag{0043}[ct][ct]{\matlabtextAB $2$}%
\psfrag{0044}[ct][ct]{\matlabtextAB $3$}%
\psfrag{0045}[ct][ct]{\matlabtextAB $4$}%
\psfrag{0046}[ct][ct]{\matlabtextAB $5$}%
\psfrag{0047}[ct][ct]{\matlabtextAB $6$}%
\psfrag{0048}[ct][ct]{\matlabtextAB $7$}%
\psfrag{0049}[ct][ct]{\matlabtextAB $8$}%
\psfrag{0050}[ct][ct]{\matlabtextAB $9$}%
\psfrag{0051}[ct][ct]{\matlabtextAB $10$}%
\psfrag{0052}[ct][ct]{\matlabtextAB $0$}%
\psfrag{0053}[ct][ct]{\matlabtextAB $1$}%
\psfrag{0054}[ct][ct]{\matlabtextAB $2$}%
\psfrag{0055}[ct][ct]{\matlabtextAB $3$}%
\psfrag{0056}[ct][ct]{\matlabtextAB $4$}%
\psfrag{0057}[ct][ct]{\matlabtextAB $5$}%
\psfrag{0058}[ct][ct]{\matlabtextAB $6$}%
\psfrag{0059}[ct][ct]{\matlabtextAB $7$}%
\psfrag{0060}[ct][ct]{\matlabtextAB $8$}%
\psfrag{0061}[ct][ct]{\matlabtextAB $9$}%
\psfrag{0062}[ct][ct]{\matlabtextAB $10$}%
\psfrag{0011}[rc][rc]{\matlabtextAB $0$}%
\psfrag{0012}[rc][rc]{\matlabtextAB $500$}%
\psfrag{0013}[rc][rc]{\matlabtextAB $1000$}%
\psfrag{0014}[rc][rc]{\matlabtextAB $1500$}%
\psfrag{0037}[rc][rc]{\matlabtextAB $0$}%
\psfrag{0038}[rc][rc]{\matlabtextAB $500$}%
\psfrag{0039}[rc][rc]{\matlabtextAB $1000$}%
\psfrag{0040}[rc][rc]{\matlabtextAB $1500$}%
\psfrag{0063}[rc][rc]{\matlabtextAB $0$}%
\psfrag{0064}[rc][rc]{\matlabtextAB $500$}%
\psfrag{0065}[rc][rc]{\matlabtextAB $1000$}%
\psfrag{0066}[rc][rc]{\matlabtextAB $1500$}%
\providecommand\matlabtextAC{\color[rgb]{0.000,0.000,0.000}\fontsize{8}{8}\selectfont\strut}%
\psfrag{0090}[bc][bc]{\matlabtextAC $r_h$}%
\psfrag{0091}[tc][tc]{\matlabtextAC $\log_{10}v$}%
\psfrag{0092}[cl][cl]{\matlabtextAC 22-02-2007}%
\psfrag{0093}[cl][cl]{\matlabtextAC 21-02-2007}%
\psfrag{0094}[cl][cl]{\matlabtextAC 20-02-2007}%
\psfrag{0095}[cl][cl]{\matlabtextAC 19-02-2007}%
\def\matlabfragNegXTick{\mathord{\makebox[0pt][r]{$-$}}}
\providecommand\matlabtextAD{\color[rgb]{0.000,0.000,0.000}\fontsize{5}{5}\selectfont\strut}%
\psfrag{0074}[ct][ct]{\matlabtextAD $2$}%
\psfrag{0075}[ct][ct]{\matlabtextAD $3$}%
\psfrag{0076}[ct][ct]{\matlabtextAD $4$}%
\psfrag{0077}[ct][ct]{\matlabtextAD $5$}%
\psfrag{0078}[ct][ct]{\matlabtextAD $6$}%
\psfrag{0079}[ct][ct]{\matlabtextAD $7$}%
\psfrag{0080}[ct][ct]{\matlabtextAD $8$}%
\psfrag{0081}[ct][ct]{\matlabtextAD $9$}%
\psfrag{0082}[rc][rc]{\matlabtextAD $4.5$}%
\psfrag{0083}[rc][rc]{\matlabtextAD $5$}%
\psfrag{0084}[rc][rc]{\matlabtextAD $5.5$}%
\psfrag{0085}[rc][rc]{\matlabtextAD $6$}%
\psfrag{0086}[rc][rc]{\matlabtextAD $6.5$}%
\psfrag{0087}[rc][rc]{\matlabtextAD $7$}%
\psfrag{0088}[rc][rc]{\matlabtextAD $7.5$}%
\psfrag{0089}[rc][rc]{\matlabtextAD $8$}%
\providecommand\matlabtextAE{\color[rgb]{0.000,0.000,0.000}\fontsize{8}{8}\selectfont\strut}%
\psfrag{0144}[tc][tc]{\matlabtextAE $\log_{10}v$}%
\psfrag{0145}[cl][cl]{\matlabtextAE 22-02-2007}%
\psfrag{0146}[cl][cl]{\matlabtextAE 21-02-2007}%
\psfrag{0147}[cl][cl]{\matlabtextAE 20-02-2007}%
\psfrag{0148}[cl][cl]{\matlabtextAE 19-02-2007}%
\def\matlabfragNegXTick{\mathord{\makebox[0pt][r]{$-$}}}
\providecommand\matlabtextAF{\color[rgb]{0.000,0.000,0.000}\fontsize{5}{5}\selectfont\strut}%
\psfrag{0096}[ct][ct]{\matlabtextAF $4$}%
\psfrag{0097}[ct][ct]{\matlabtextAF $5$}%
\psfrag{0098}[ct][ct]{\matlabtextAF $6$}%
\psfrag{0099}[ct][ct]{\matlabtextAF $7$}%
\psfrag{0100}[ct][ct]{\matlabtextAF $8$}%
\psfrag{0101}[ct][ct]{\matlabtextAF $9$}%
\psfrag{0102}[ct][ct]{\matlabtextAF $10$}%
\psfrag{0108}[ct][ct]{\matlabtextAF $4$}%
\psfrag{0109}[ct][ct]{\matlabtextAF $5$}%
\psfrag{0110}[ct][ct]{\matlabtextAF $6$}%
\psfrag{0111}[ct][ct]{\matlabtextAF $7$}%
\psfrag{0112}[ct][ct]{\matlabtextAF $8$}%
\psfrag{0113}[ct][ct]{\matlabtextAF $9$}%
\psfrag{0114}[ct][ct]{\matlabtextAF $10$}%
\psfrag{0120}[ct][ct]{\matlabtextAF $4$}%
\psfrag{0121}[ct][ct]{\matlabtextAF $5$}%
\psfrag{0122}[ct][ct]{\matlabtextAF $6$}%
\psfrag{0123}[ct][ct]{\matlabtextAF $7$}%
\psfrag{0124}[ct][ct]{\matlabtextAF $8$}%
\psfrag{0125}[ct][ct]{\matlabtextAF $9$}%
\psfrag{0126}[ct][ct]{\matlabtextAF $10$}%
\psfrag{0132}[ct][ct]{\matlabtextAF $4$}%
\psfrag{0133}[ct][ct]{\matlabtextAF $5$}%
\psfrag{0134}[ct][ct]{\matlabtextAF $6$}%
\psfrag{0135}[ct][ct]{\matlabtextAF $7$}%
\psfrag{0136}[ct][ct]{\matlabtextAF $8$}%
\psfrag{0137}[ct][ct]{\matlabtextAF $9$}%
\psfrag{0138}[ct][ct]{\matlabtextAF $10$}%
\psfrag{0103}[rc][rc]{\matlabtextAF $0$}%
\psfrag{0104}[rc][rc]{\matlabtextAF $10$}%
\psfrag{0105}[rc][rc]{\matlabtextAF $20$}%
\psfrag{0106}[rc][rc]{\matlabtextAF $30$}%
\psfrag{0107}[rc][rc]{\matlabtextAF $40$}%
\psfrag{0115}[rc][rc]{\matlabtextAF $0$}%
\psfrag{0116}[rc][rc]{\matlabtextAF $10$}%
\psfrag{0117}[rc][rc]{\matlabtextAF $20$}%
\psfrag{0118}[rc][rc]{\matlabtextAF $30$}%
\psfrag{0119}[rc][rc]{\matlabtextAF $40$}%
\psfrag{0127}[rc][rc]{\matlabtextAF $0$}%
\psfrag{0128}[rc][rc]{\matlabtextAF $10$}%
\psfrag{0129}[rc][rc]{\matlabtextAF $20$}%
\psfrag{0130}[rc][rc]{\matlabtextAF $30$}%
\psfrag{0131}[rc][rc]{\matlabtextAF $40$}%
\psfrag{0139}[rc][rc]{\matlabtextAF $0$}%
\psfrag{0140}[rc][rc]{\matlabtextAF $10$}%
\psfrag{0141}[rc][rc]{\matlabtextAF $20$}%
\psfrag{0142}[rc][rc]{\matlabtextAF $30$}%
\psfrag{0143}[rc][rc]{\matlabtextAF $40$}%
\providecommand\matlabtextAG{\color[rgb]{0.000,0.000,0.000}\fontsize{8}{8}\selectfont\strut}%
\psfrag{0185}[tc][tc]{\matlabtextAG $r_h$}%
\psfrag{0186}[cl][cl]{\matlabtextAG 22-02-2007}%
\psfrag{0187}[cl][cl]{\matlabtextAG 21-02-2007}%
\psfrag{0188}[cl][cl]{\matlabtextAG 20-02-2007}%
\psfrag{0189}[cl][cl]{\matlabtextAG 19-02-2007}%
\def\matlabfragNegXTick{\mathord{\makebox[0pt][r]{$-$}}}
\providecommand\matlabtextAH{\color[rgb]{0.000,0.000,0.000}\fontsize{5}{5}\selectfont\strut}%
\psfrag{0149}[ct][ct]{\matlabtextAH $5.75$}%
\psfrag{0150}[ct][ct]{\matlabtextAH $6$}%
\psfrag{0151}[ct][ct]{\matlabtextAH $6.25$}%
\psfrag{0152}[ct][ct]{\matlabtextAH $6.5$}%
\psfrag{0153}[ct][ct]{\matlabtextAH $6.75$}%
\psfrag{0158}[ct][ct]{\matlabtextAH $5.75$}%
\psfrag{0159}[ct][ct]{\matlabtextAH $6$}%
\psfrag{0160}[ct][ct]{\matlabtextAH $6.25$}%
\psfrag{0161}[ct][ct]{\matlabtextAH $6.5$}%
\psfrag{0162}[ct][ct]{\matlabtextAH $6.75$}%
\psfrag{0167}[ct][ct]{\matlabtextAH $5.75$}%
\psfrag{0168}[ct][ct]{\matlabtextAH $6$}%
\psfrag{0169}[ct][ct]{\matlabtextAH $6.25$}%
\psfrag{0170}[ct][ct]{\matlabtextAH $6.5$}%
\psfrag{0171}[ct][ct]{\matlabtextAH $6.75$}%
\psfrag{0176}[ct][ct]{\matlabtextAH $5.75$}%
\psfrag{0177}[ct][ct]{\matlabtextAH $6$}%
\psfrag{0178}[ct][ct]{\matlabtextAH $6.25$}%
\psfrag{0179}[ct][ct]{\matlabtextAH $6.5$}%
\psfrag{0180}[ct][ct]{\matlabtextAH $6.75$}%
\psfrag{0154}[rc][rc]{\matlabtextAH $0$}%
\psfrag{0155}[rc][rc]{\matlabtextAH $20$}%
\psfrag{0156}[rc][rc]{\matlabtextAH $40$}%
\psfrag{0157}[rc][rc]{\matlabtextAH $60$}%
\psfrag{0163}[rc][rc]{\matlabtextAH $0$}%
\psfrag{0164}[rc][rc]{\matlabtextAH $20$}%
\psfrag{0165}[rc][rc]{\matlabtextAH $40$}%
\psfrag{0166}[rc][rc]{\matlabtextAH $60$}%
\psfrag{0172}[rc][rc]{\matlabtextAH $0$}%
\psfrag{0173}[rc][rc]{\matlabtextAH $20$}%
\psfrag{0174}[rc][rc]{\matlabtextAH $40$}%
\psfrag{0175}[rc][rc]{\matlabtextAH $60$}%
\psfrag{0181}[rc][rc]{\matlabtextAH $0$}%
\psfrag{0182}[rc][rc]{\matlabtextAH $20$}%
\psfrag{0183}[rc][rc]{\matlabtextAH $40$}%
\psfrag{0184}[rc][rc]{\matlabtextAH $60$}%
\providecommand\matlabtextAI{\color[rgb]{0.000,0.000,0.000}\fontsize{5}{5}\selectfont\strut}%
\psfrag{0254}[bl][bl]{\matlabtextAI W/34/+0.36}%
\psfrag{0255}[bl][bl]{\matlabtextAI W/20/-0.36}%
\psfrag{0256}[bl][bl]{\matlabtextAI W/27/+0.08}%
\psfrag{0257}[bl][bl]{\matlabtextAI W/ 7/-0.09}%
\psfrag{0258}[bl][bl]{\matlabtextAI T/30/-0.21}%
\psfrag{0259}[bl][bl]{\matlabtextAI T/21/-0.62}%
\psfrag{0260}[bl][bl]{\matlabtextAI T/29/-0.75}%
\psfrag{0261}[bl][bl]{\matlabtextAI T/15/-0.76}%
\psfrag{0262}[bl][bl]{\matlabtextAI BA/30/-1.53}%
\psfrag{0263}[bl][bl]{\matlabtextAI BA/30/-0.32}%
\psfrag{0264}[bl][bl]{\matlabtextAI BA/33/-0.10}%
\psfrag{0265}[bl][bl]{\matlabtextAI BA/23/+0.03}%
\psfrag{0266}[bl][bl]{\matlabtextAI AV/51/+1.08}%
\psfrag{0267}[bl][bl]{\matlabtextAI AV/37/+0.55}%
\psfrag{0268}[bl][bl]{\matlabtextAI AV/51/+1.39}%
\psfrag{0269}[bl][bl]{\matlabtextAI AV/20/-0.32}%
\psfrag{0270}[bl][bl]{\matlabtextAI BP/57/+0.16}%
\psfrag{0271}[bl][bl]{\matlabtextAI BP/41/+1.38}%
\psfrag{0272}[bl][bl]{\matlabtextAI BP/57/+0.80}%
\psfrag{0273}[bl][bl]{\matlabtextAI BP/17/+2.08}%
\psfrag{0274}[bl][bl]{\matlabtextAI D/66/-1.25}%
\psfrag{0275}[bl][bl]{\matlabtextAI D/51/-0.76}%
\psfrag{0276}[bl][bl]{\matlabtextAI D/60/-0.28}%
\psfrag{0277}[bl][bl]{\matlabtextAI D/15/-0.51}%
\def\matlabfragNegXTick{\mathord{\makebox[0pt][r]{$-$}}}
\psfrag{0190}[ct][ct]{\matlabtextAI 0}%
\psfrag{0191}[ct][ct]{\matlabtextAI 1}%
\psfrag{0192}[ct][ct]{\matlabtextAI 2}%
\psfrag{0193}[ct][ct]{\matlabtextAI 3}%
\psfrag{0194}[ct][ct]{\matlabtextAI 4}%
\psfrag{0195}[ct][ct]{\matlabtextAI 5}%
\psfrag{0196}[ct][ct]{\matlabtextAI 6}%
\psfrag{0197}[ct][ct]{\matlabtextAI 7}%
\psfrag{0198}[ct][ct]{\matlabtextAI 8}%
\psfrag{0199}[ct][ct]{\matlabtextAI 9}%
\psfrag{0200}[ct][ct]{\matlabtextAI }%
\psfrag{0201}[ct][ct]{\matlabtextAI 0}%
\psfrag{0202}[ct][ct]{\matlabtextAI 1}%
\psfrag{0203}[ct][ct]{\matlabtextAI 2}%
\psfrag{0204}[ct][ct]{\matlabtextAI 3}%
\psfrag{0205}[ct][ct]{\matlabtextAI 4}%
\psfrag{0206}[ct][ct]{\matlabtextAI 5}%
\psfrag{0207}[ct][ct]{\matlabtextAI 6}%
\psfrag{0208}[ct][ct]{\matlabtextAI 7}%
\psfrag{0209}[ct][ct]{\matlabtextAI 8}%
\psfrag{0210}[ct][ct]{\matlabtextAI 9}%
\psfrag{0211}[ct][ct]{\matlabtextAI }%
\psfrag{0212}[ct][ct]{\matlabtextAI 0}%
\psfrag{0213}[ct][ct]{\matlabtextAI 1}%
\psfrag{0214}[ct][ct]{\matlabtextAI 2}%
\psfrag{0215}[ct][ct]{\matlabtextAI 3}%
\psfrag{0216}[ct][ct]{\matlabtextAI 4}%
\psfrag{0217}[ct][ct]{\matlabtextAI 5}%
\psfrag{0218}[ct][ct]{\matlabtextAI 6}%
\psfrag{0219}[ct][ct]{\matlabtextAI 7}%
\psfrag{0220}[ct][ct]{\matlabtextAI 8}%
\psfrag{0221}[ct][ct]{\matlabtextAI 9}%
\psfrag{0222}[ct][ct]{\matlabtextAI }%
\psfrag{0223}[ct][ct]{\matlabtextAI 0}%
\psfrag{0224}[ct][ct]{\matlabtextAI 1}%
\psfrag{0225}[ct][ct]{\matlabtextAI 2}%
\psfrag{0226}[ct][ct]{\matlabtextAI 3}%
\psfrag{0227}[ct][ct]{\matlabtextAI 4}%
\psfrag{0228}[ct][ct]{\matlabtextAI 5}%
\psfrag{0229}[ct][ct]{\matlabtextAI 6}%
\psfrag{0230}[ct][ct]{\matlabtextAI 7}%
\psfrag{0231}[ct][ct]{\matlabtextAI 8}%
\psfrag{0232}[ct][ct]{\matlabtextAI 9}%
\psfrag{0233}[ct][ct]{\matlabtextAI }%
\psfrag{0234}[rc][rc]{\matlabtextAI $0$}%
\psfrag{0235}[rc][rc]{\matlabtextAI $50$}%
\psfrag{0236}[rc][rc]{\matlabtextAI $100$}%
\psfrag{0237}[rc][rc]{\matlabtextAI $150$}%
\psfrag{0238}[rc][rc]{\matlabtextAI $0$}%
\psfrag{0239}[rc][rc]{\matlabtextAI $100$}%
\psfrag{0240}[rc][rc]{\matlabtextAI $200$}%
\psfrag{0241}[rc][rc]{\matlabtextAI $0$}%
\psfrag{0242}[rc][rc]{\matlabtextAI $20$}%
\psfrag{0243}[rc][rc]{\matlabtextAI $40$}%
\psfrag{0244}[rc][rc]{\matlabtextAI $0$}%
\psfrag{0245}[rc][rc]{\matlabtextAI $100$}%
\psfrag{0246}[rc][rc]{\matlabtextAI $200$}%
\psfrag{0247}[rc][rc]{\matlabtextAI $0$}%
\psfrag{0248}[rc][rc]{\matlabtextAI $100$}%
\psfrag{0249}[rc][rc]{\matlabtextAI $200$}%
\psfrag{0250}[rc][rc]{\matlabtextAI $0$}%
\psfrag{0251}[rc][rc]{\matlabtextAI $100$}%
\psfrag{0252}[rc][rc]{\matlabtextAI $200$}%
\psfrag{0253}[rc][rc]{\matlabtextAI $300$}%
\providecommand\matlabtextAK{\color[rgb]{0.000,0.000,0.000}\fontsize{8}{8}\selectfont\strut}%
\psfrag{0391}[bc][bc]{\matlabtextAK $\log_{10}|\Delta l|$}%
\psfrag{0392}[tc][tc]{\matlabtextAK $\log_{10}v$}%
\psfrag{0393}[bc][bc]{\matlabtextAK $\Delta l$}%
\psfrag{0394}[tc][tc]{\matlabtextAK $\Delta v$}%
\def\matlabfragNegXTick{\mathord{\makebox[0pt][r]{$-$}}}
\providecommand\matlabtextAL{\color[rgb]{0.000,0.000,0.000}\fontsize{5}{5}\selectfont\strut}%
\psfrag{0366}[ct][ct]{\matlabtextAL $6$}%
\psfrag{0367}[ct][ct]{\matlabtextAL $7$}%
\psfrag{0368}[ct][ct]{\matlabtextAL $8$}%
\psfrag{0369}[ct][ct]{\matlabtextAL $9$}%
\psfrag{0370}[ct][ct]{\matlabtextAL $10$}%
\psfrag{0371}[ct][ct]{\matlabtextAL $11$}%
\psfrag{0381}[ct][ct]{\matlabtextAL $\matlabfragNegXTick 2$}%
\psfrag{0382}[ct][ct]{\matlabtextAL $\matlabfragNegXTick 1$}%
\psfrag{0383}[ct][ct]{\matlabtextAL $0$}%
\psfrag{0384}[ct][ct]{\matlabtextAL $1$}%
\psfrag{0385}[ct][ct]{\matlabtextAL $2$}%
\psfrag{0372}[rc][rc]{\matlabtextAL $5.5$}%
\psfrag{0373}[rc][rc]{\matlabtextAL $6$}%
\psfrag{0374}[rc][rc]{\matlabtextAL $6.5$}%
\psfrag{0375}[rc][rc]{\matlabtextAL $7$}%
\psfrag{0376}[rc][rc]{\matlabtextAL $7.5$}%
\psfrag{0377}[rc][rc]{\matlabtextAL $8$}%
\psfrag{0378}[rc][rc]{\matlabtextAL $8.5$}%
\psfrag{0379}[rc][rc]{\matlabtextAL $9$}%
\psfrag{0380}[rc][rc]{\matlabtextAL $9.5$}%
\psfrag{0386}[rc][rc]{\matlabtextAL $-2$}%
\psfrag{0387}[rc][rc]{\matlabtextAL $-1$}%
\psfrag{0388}[rc][rc]{\matlabtextAL $0$}%
\psfrag{0389}[rc][rc]{\matlabtextAL $1$}%
\psfrag{0390}[rc][rc]{\matlabtextAL $2$}%
\providecommand\matlabtextAM{\color[rgb]{0.000,0.000,0.000}\fontsize{8}{8}\selectfont\strut}%
\psfrag{0461}[cc][cc]{\matlabtextAM $\log_{10}v$ on 22-02-2007}%
\psfrag{0462}[tc][tc]{\matlabtextAM $\log_{10}v$ on 21-02-2007}%
\psfrag{0463}[cc][cc]{\matlabtextAM $\log_{10}v$ on 21-02-2007}%
\psfrag{0464}[tc][tc]{\matlabtextAM $\log_{10}v$ on 20-02-2007}%
\psfrag{0465}[cc][cc]{\matlabtextAM $\log_{10}v$ on 20-02-2007}%
\psfrag{0466}[tc][tc]{\matlabtextAM $\log_{10}v$ on 19-02-2007}%
\def\matlabfragNegXTick{\mathord{\makebox[0pt][r]{$-$}}}
\providecommand\matlabtextAN{\color[rgb]{0.000,0.000,0.000}\fontsize{5}{5}\selectfont\strut}%
\psfrag{0395}[ct][ct]{\matlabtextAN $0$}%
\psfrag{0396}[ct][ct]{\matlabtextAN $1$}%
\psfrag{0397}[ct][ct]{\matlabtextAN $2$}%
\psfrag{0398}[ct][ct]{\matlabtextAN $3$}%
\psfrag{0399}[ct][ct]{\matlabtextAN $4$}%
\psfrag{0400}[ct][ct]{\matlabtextAN $5$}%
\psfrag{0401}[ct][ct]{\matlabtextAN $6$}%
\psfrag{0402}[ct][ct]{\matlabtextAN $7$}%
\psfrag{0403}[ct][ct]{\matlabtextAN $8$}%
\psfrag{0404}[ct][ct]{\matlabtextAN $9$}%
\psfrag{0405}[ct][ct]{\matlabtextAN $10$}%
\psfrag{0417}[ct][ct]{\matlabtextAN $0$}%
\psfrag{0418}[ct][ct]{\matlabtextAN $1$}%
\psfrag{0419}[ct][ct]{\matlabtextAN $2$}%
\psfrag{0420}[ct][ct]{\matlabtextAN $3$}%
\psfrag{0421}[ct][ct]{\matlabtextAN $4$}%
\psfrag{0422}[ct][ct]{\matlabtextAN $5$}%
\psfrag{0423}[ct][ct]{\matlabtextAN $6$}%
\psfrag{0424}[ct][ct]{\matlabtextAN $7$}%
\psfrag{0425}[ct][ct]{\matlabtextAN $8$}%
\psfrag{0426}[ct][ct]{\matlabtextAN $9$}%
\psfrag{0427}[ct][ct]{\matlabtextAN $10$}%
\psfrag{0439}[ct][ct]{\matlabtextAN $0$}%
\psfrag{0440}[ct][ct]{\matlabtextAN $1$}%
\psfrag{0441}[ct][ct]{\matlabtextAN $2$}%
\psfrag{0442}[ct][ct]{\matlabtextAN $3$}%
\psfrag{0443}[ct][ct]{\matlabtextAN $4$}%
\psfrag{0444}[ct][ct]{\matlabtextAN $5$}%
\psfrag{0445}[ct][ct]{\matlabtextAN $6$}%
\psfrag{0446}[ct][ct]{\matlabtextAN $7$}%
\psfrag{0447}[ct][ct]{\matlabtextAN $8$}%
\psfrag{0448}[ct][ct]{\matlabtextAN $9$}%
\psfrag{0449}[ct][ct]{\matlabtextAN $10$}%
\psfrag{0406}[rc][rc]{\matlabtextAN $0$}%
\psfrag{0407}[rc][rc]{\matlabtextAN $1$}%
\psfrag{0408}[rc][rc]{\matlabtextAN $2$}%
\psfrag{0409}[rc][rc]{\matlabtextAN $3$}%
\psfrag{0410}[rc][rc]{\matlabtextAN $4$}%
\psfrag{0411}[rc][rc]{\matlabtextAN $5$}%
\psfrag{0412}[rc][rc]{\matlabtextAN $6$}%
\psfrag{0413}[rc][rc]{\matlabtextAN $7$}%
\psfrag{0414}[rc][rc]{\matlabtextAN $8$}%
\psfrag{0415}[rc][rc]{\matlabtextAN $9$}%
\psfrag{0416}[rc][rc]{\matlabtextAN $10$}%
\psfrag{0428}[rc][rc]{\matlabtextAN $0$}%
\psfrag{0429}[rc][rc]{\matlabtextAN $1$}%
\psfrag{0430}[rc][rc]{\matlabtextAN $2$}%
\psfrag{0431}[rc][rc]{\matlabtextAN $3$}%
\psfrag{0432}[rc][rc]{\matlabtextAN $4$}%
\psfrag{0433}[rc][rc]{\matlabtextAN $5$}%
\psfrag{0434}[rc][rc]{\matlabtextAN $6$}%
\psfrag{0435}[rc][rc]{\matlabtextAN $7$}%
\psfrag{0436}[rc][rc]{\matlabtextAN $8$}%
\psfrag{0437}[rc][rc]{\matlabtextAN $9$}%
\psfrag{0438}[rc][rc]{\matlabtextAN $10$}%
\psfrag{0450}[rc][rc]{\matlabtextAN $0$}%
\psfrag{0451}[rc][rc]{\matlabtextAN $1$}%
\psfrag{0452}[rc][rc]{\matlabtextAN $2$}%
\psfrag{0453}[rc][rc]{\matlabtextAN $3$}%
\psfrag{0454}[rc][rc]{\matlabtextAN $4$}%
\psfrag{0455}[rc][rc]{\matlabtextAN $5$}%
\psfrag{0456}[rc][rc]{\matlabtextAN $6$}%
\psfrag{0457}[rc][rc]{\matlabtextAN $7$}%
\psfrag{0458}[rc][rc]{\matlabtextAN $8$}%
\psfrag{0459}[rc][rc]{\matlabtextAN $9$}%
\psfrag{0460}[rc][rc]{\matlabtextAN $10$}%
\providecommand\matlabtextAO{\color[rgb]{0.000,0.000,0.000}\fontsize{8}{8}\selectfont\strut}%
\psfrag{0503}[cc][cc]{\matlabtextAO $\log_{10}v$ on 22-02-2007}%
\psfrag{0504}[tc][tc]{\matlabtextAO $\log_{10}v$ on 21-02-2007}%
\psfrag{0505}[cc][cc]{\matlabtextAO $\log_{10}v$ on 21-02-2007}%
\psfrag{0506}[tc][tc]{\matlabtextAO $\log_{10}v$ on 20-02-2007}%
\psfrag{0507}[cc][cc]{\matlabtextAO $\log_{10}v$ on 20-02-2007}%
\psfrag{0508}[tc][tc]{\matlabtextAO $\log_{10}v$ on 19-02-2007}%
\def\matlabfragNegXTick{\mathord{\makebox[0pt][r]{$-$}}}
\providecommand\matlabtextAP{\color[rgb]{0.000,0.000,0.000}\fontsize{5}{5}\selectfont\strut}%
\psfrag{0467}[ct][ct]{\matlabtextAP $5$}%
\psfrag{0468}[ct][ct]{\matlabtextAP $6$}%
\psfrag{0469}[ct][ct]{\matlabtextAP $7$}%
\psfrag{0470}[ct][ct]{\matlabtextAP $8$}%
\psfrag{0471}[ct][ct]{\matlabtextAP $9$}%
\psfrag{0472}[ct][ct]{\matlabtextAP $10$}%
\psfrag{0479}[ct][ct]{\matlabtextAP $5$}%
\psfrag{0480}[ct][ct]{\matlabtextAP $6$}%
\psfrag{0481}[ct][ct]{\matlabtextAP $7$}%
\psfrag{0482}[ct][ct]{\matlabtextAP $8$}%
\psfrag{0483}[ct][ct]{\matlabtextAP $9$}%
\psfrag{0484}[ct][ct]{\matlabtextAP $10$}%
\psfrag{0491}[ct][ct]{\matlabtextAP $5$}%
\psfrag{0492}[ct][ct]{\matlabtextAP $6$}%
\psfrag{0493}[ct][ct]{\matlabtextAP $7$}%
\psfrag{0494}[ct][ct]{\matlabtextAP $8$}%
\psfrag{0495}[ct][ct]{\matlabtextAP $9$}%
\psfrag{0496}[ct][ct]{\matlabtextAP $10$}%
\psfrag{0473}[rc][rc]{\matlabtextAP $5$}%
\psfrag{0474}[rc][rc]{\matlabtextAP $6$}%
\psfrag{0475}[rc][rc]{\matlabtextAP $7$}%
\psfrag{0476}[rc][rc]{\matlabtextAP $8$}%
\psfrag{0477}[rc][rc]{\matlabtextAP $9$}%
\psfrag{0478}[rc][rc]{\matlabtextAP $10$}%
\psfrag{0485}[rc][rc]{\matlabtextAP $5$}%
\psfrag{0486}[rc][rc]{\matlabtextAP $6$}%
\psfrag{0487}[rc][rc]{\matlabtextAP $7$}%
\psfrag{0488}[rc][rc]{\matlabtextAP $8$}%
\psfrag{0489}[rc][rc]{\matlabtextAP $9$}%
\psfrag{0490}[rc][rc]{\matlabtextAP $10$}%
\psfrag{0497}[rc][rc]{\matlabtextAP $5$}%
\psfrag{0498}[rc][rc]{\matlabtextAP $6$}%
\psfrag{0499}[rc][rc]{\matlabtextAP $7$}%
\psfrag{0500}[rc][rc]{\matlabtextAP $8$}%
\psfrag{0501}[rc][rc]{\matlabtextAP $9$}%
\psfrag{0502}[rc][rc]{\matlabtextAP $10$}%
\providecommand\matlabtextBA{\color[rgb]{0.000,0.000,0.000}\fontsize{8}{8}\selectfont\strut}%
\psfrag{0591}[cc][cc]{\matlabtextBA $\log_{10}l$}%
\psfrag{0592}[tc][tc]{\matlabtextBA $\log_{10}v$}%
\def\matlabfragNegXTick{\mathord{\makebox[0pt][r]{$-$}}}
\providecommand\matlabtextBB{\color[rgb]{0.000,0.000,0.000}\fontsize{5}{5}\selectfont\strut}%
\psfrag{0579}[ct][ct]{\matlabtextBB $5$}%
\psfrag{0580}[ct][ct]{\matlabtextBB $6$}%
\psfrag{0581}[ct][ct]{\matlabtextBB $7$}%
\psfrag{0582}[ct][ct]{\matlabtextBB $8$}%
\psfrag{0583}[ct][ct]{\matlabtextBB $9$}%
\psfrag{0584}[ct][ct]{\matlabtextBB $10$}%
\psfrag{0585}[rc][rc]{\matlabtextBB $5$}%
\psfrag{0586}[rc][rc]{\matlabtextBB $6$}%
\psfrag{0587}[rc][rc]{\matlabtextBB $7$}%
\psfrag{0588}[rc][rc]{\matlabtextBB $8$}%
\psfrag{0589}[rc][rc]{\matlabtextBB $9$}%
\psfrag{0590}[rc][rc]{\matlabtextBB $10$}%
\providecommand\matlabtextBO{\color[rgb]{0.000,0.000,0.000}\fontsize{8}{8}\selectfont\strut}%
\psfrag{0848}[tc][tc]{\matlabtextBO $\log_{10}v$}%
\psfrag{0849}[bl][bl]{\matlabtextBO 22-02-2007}%
\psfrag{0850}[tc][tc]{\matlabtextBO $\log_{10}v$}%
\psfrag{0851}[bl][bl]{\matlabtextBO 21-02-2007}%
\psfrag{0852}[bl][bl]{\matlabtextBO 20-02-2007}%
\psfrag{0853}[bl][bl]{\matlabtextBO 19-02-2007}%
\def\matlabfragNegXTick{\mathord{\makebox[0pt][r]{$-$}}}
\providecommand\matlabtextBP{\color[rgb]{0.000,0.000,0.000}\fontsize{5}{5}\selectfont\strut}%
\psfrag{0812}[ct][ct]{\matlabtextBP $0$}%
\psfrag{0813}[ct][ct]{\matlabtextBP $1$}%
\psfrag{0814}[ct][ct]{\matlabtextBP $2$}%
\psfrag{0815}[ct][ct]{\matlabtextBP $3$}%
\psfrag{0816}[ct][ct]{\matlabtextBP $4$}%
\psfrag{0817}[ct][ct]{\matlabtextBP $5$}%
\psfrag{0818}[ct][ct]{\matlabtextBP $6$}%
\psfrag{0819}[ct][ct]{\matlabtextBP $7$}%
\psfrag{0820}[ct][ct]{\matlabtextBP $8$}%
\psfrag{0821}[ct][ct]{\matlabtextBP $9$}%
\psfrag{0822}[ct][ct]{\matlabtextBP $10$}%
\psfrag{0823}[ct][ct]{\matlabtextBP $0$}%
\psfrag{0824}[ct][ct]{\matlabtextBP $1$}%
\psfrag{0825}[ct][ct]{\matlabtextBP $2$}%
\psfrag{0826}[ct][ct]{\matlabtextBP $3$}%
\psfrag{0827}[ct][ct]{\matlabtextBP $4$}%
\psfrag{0828}[ct][ct]{\matlabtextBP $5$}%
\psfrag{0829}[ct][ct]{\matlabtextBP $6$}%
\psfrag{0830}[ct][ct]{\matlabtextBP $7$}%
\psfrag{0831}[ct][ct]{\matlabtextBP $8$}%
\psfrag{0832}[ct][ct]{\matlabtextBP $9$}%
\psfrag{0833}[ct][ct]{\matlabtextBP $10$}%
\psfrag{0834}[rc][rc]{\matlabtextBP $0$}%
\psfrag{0835}[rc][rc]{\matlabtextBP $5$}%
\psfrag{0836}[rc][rc]{\matlabtextBP $10$}%
\psfrag{0837}[rc][rc]{\matlabtextBP $15$}%
\psfrag{0838}[rc][rc]{\matlabtextBP $20$}%
\psfrag{0839}[rc][rc]{\matlabtextBP $25$}%
\psfrag{0840}[rc][rc]{\matlabtextBP $30$}%
\psfrag{0841}[rc][rc]{\matlabtextBP $0$}%
\psfrag{0842}[rc][rc]{\matlabtextBP $5$}%
\psfrag{0843}[rc][rc]{\matlabtextBP $10$}%
\psfrag{0844}[rc][rc]{\matlabtextBP $15$}%
\psfrag{0845}[rc][rc]{\matlabtextBP $20$}%
\psfrag{0846}[rc][rc]{\matlabtextBP $25$}%
\psfrag{0847}[rc][rc]{\matlabtextBP $30$}%
\providecommand\matlabtextBE{\color[rgb]{0.000,0.000,0.000}\fontsize{8}{8}\selectfont\strut}%
\psfrag{0706}[tc][tc]{\matlabtextBE $d$}%
\psfrag{0707}[cl][cl]{\matlabtextBE 22-02-2007}%
\psfrag{0708}[cl][cl]{\matlabtextBE 21-02-2007}%
\psfrag{0709}[cl][cl]{\matlabtextBE 20-02-2007}%
\psfrag{0710}[cl][cl]{\matlabtextBE 19-02-2007}%
\def\matlabfragNegXTick{\mathord{\makebox[0pt][r]{$-$}}}
\providecommand\matlabtextBF{\color[rgb]{0.000,0.000,0.000}\fontsize{5}{5}\selectfont\strut}%
\psfrag{0677}[ct][ct]{\matlabtextBF $0$}%
\psfrag{0678}[ct][ct]{\matlabtextBF $5$}%
\psfrag{0679}[ct][ct]{\matlabtextBF $10$}%
\psfrag{0680}[ct][ct]{\matlabtextBF $15$}%
\psfrag{0681}[ct][ct]{\matlabtextBF $20$}%
\psfrag{0682}[ct][ct]{\matlabtextBF $25$}%
\psfrag{0683}[ct][ct]{\matlabtextBF $30$}%
\psfrag{0684}[ct][ct]{\matlabtextBF $35$}%
\psfrag{0685}[ct][ct]{\matlabtextBF $40$}%
\psfrag{0657}[lc][lc]{\matlabtextBF $9$}%
\psfrag{0658}[lc][lc]{\matlabtextBF $5$}%
\psfrag{0659}[lc][lc]{\matlabtextBF $0$}%
\psfrag{0660}[lc][lc]{\matlabtextBF $5$}%
\psfrag{0661}[lc][lc]{\matlabtextBF $9$}%
\psfrag{0662}[lc][lc]{\matlabtextBF $9$}%
\psfrag{0663}[lc][lc]{\matlabtextBF $5$}%
\psfrag{0664}[lc][lc]{\matlabtextBF $0$}%
\psfrag{0665}[lc][lc]{\matlabtextBF $5$}%
\psfrag{0666}[lc][lc]{\matlabtextBF $9$}%
\psfrag{0667}[lc][lc]{\matlabtextBF $9$}%
\psfrag{0668}[lc][lc]{\matlabtextBF $5$}%
\psfrag{0669}[lc][lc]{\matlabtextBF $0$}%
\psfrag{0670}[lc][lc]{\matlabtextBF $5$}%
\psfrag{0671}[lc][lc]{\matlabtextBF $9$}%
\psfrag{0672}[lc][lc]{\matlabtextBF $9$}%
\psfrag{0673}[lc][lc]{\matlabtextBF $5$}%
\psfrag{0674}[lc][lc]{\matlabtextBF $0$}%
\psfrag{0675}[lc][lc]{\matlabtextBF $5$}%
\psfrag{0676}[lc][lc]{\matlabtextBF $9$}%
\psfrag{0686}[rc][rc]{\matlabtextBF $-9$}%
\psfrag{0687}[rc][rc]{\matlabtextBF $-5$}%
\psfrag{0688}[rc][rc]{\matlabtextBF $0$}%
\psfrag{0689}[rc][rc]{\matlabtextBF $5$}%
\psfrag{0690}[rc][rc]{\matlabtextBF $9$}%
\psfrag{0691}[rc][rc]{\matlabtextBF $-9$}%
\psfrag{0692}[rc][rc]{\matlabtextBF $-5$}%
\psfrag{0693}[rc][rc]{\matlabtextBF $0$}%
\psfrag{0694}[rc][rc]{\matlabtextBF $5$}%
\psfrag{0695}[rc][rc]{\matlabtextBF $9$}%
\psfrag{0696}[rc][rc]{\matlabtextBF $-9$}%
\psfrag{0697}[rc][rc]{\matlabtextBF $-5$}%
\psfrag{0698}[rc][rc]{\matlabtextBF $0$}%
\psfrag{0699}[rc][rc]{\matlabtextBF $5$}%
\psfrag{0700}[rc][rc]{\matlabtextBF $9$}%
\psfrag{0701}[rc][rc]{\matlabtextBF $-9$}%
\psfrag{0702}[rc][rc]{\matlabtextBF $-5$}%
\psfrag{0703}[rc][rc]{\matlabtextBF $0$}%
\psfrag{0704}[rc][rc]{\matlabtextBF $5$}%
\psfrag{0705}[rc][rc]{\matlabtextBF $9$}%
\providecommand\matlabtextBG{\color[rgb]{0.000,0.000,0.000}\fontsize{8}{8}\selectfont\strut}%
\psfrag{0715}[tc][tc]{\matlabtextBG $d$}%
\def\matlabfragNegXTick{\mathord{\makebox[0pt][r]{$-$}}}
\providecommand\matlabtextBH{\color[rgb]{0.000,0.000,0.000}\fontsize{5}{5}\selectfont\strut}%
\psfrag{0711}[ct][ct]{\matlabtextBH $10^{0}$}%
\psfrag{0712}[ct][ct]{\matlabtextBH $10^{1}$}%
\psfrag{0713}[rc][rc]{\matlabtextBH $10^{0}$}%
\psfrag{0714}[rc][rc]{\matlabtextBH $10^{1}$}%
\providecommand\matlabtextBI{\color[rgb]{0.000,0.000,0.000}\fontsize{8}{8}\selectfont\strut}%
\psfrag{0760}[tc][tc]{\matlabtextBI $d$}%
\psfrag{0761}[cl][cl]{\matlabtextBI 22-02-2007}%
\psfrag{0762}[cl][cl]{\matlabtextBI 21-02-2007}%
\psfrag{0763}[cl][cl]{\matlabtextBI 20-02-2007}%
\psfrag{0764}[cl][cl]{\matlabtextBI 19-02-2007}%
\def\matlabfragNegXTick{\mathord{\makebox[0pt][r]{$-$}}}
\providecommand\matlabtextBJ{\color[rgb]{0.000,0.000,0.000}\fontsize{5}{5}\selectfont\strut}%
\psfrag{0736}[ct][ct]{\matlabtextBJ $0$}%
\psfrag{0737}[ct][ct]{\matlabtextBJ $5$}%
\psfrag{0738}[ct][ct]{\matlabtextBJ $10$}%
\psfrag{0739}[ct][ct]{\matlabtextBJ $15$}%
\psfrag{0716}[lc][lc]{\matlabtextBJ $9$}%
\psfrag{0717}[lc][lc]{\matlabtextBJ $5$}%
\psfrag{0718}[lc][lc]{\matlabtextBJ $0$}%
\psfrag{0719}[lc][lc]{\matlabtextBJ $5$}%
\psfrag{0720}[lc][lc]{\matlabtextBJ $9$}%
\psfrag{0721}[lc][lc]{\matlabtextBJ $9$}%
\psfrag{0722}[lc][lc]{\matlabtextBJ $5$}%
\psfrag{0723}[lc][lc]{\matlabtextBJ $0$}%
\psfrag{0724}[lc][lc]{\matlabtextBJ $5$}%
\psfrag{0725}[lc][lc]{\matlabtextBJ $9$}%
\psfrag{0726}[lc][lc]{\matlabtextBJ $9$}%
\psfrag{0727}[lc][lc]{\matlabtextBJ $5$}%
\psfrag{0728}[lc][lc]{\matlabtextBJ $0$}%
\psfrag{0729}[lc][lc]{\matlabtextBJ $5$}%
\psfrag{0730}[lc][lc]{\matlabtextBJ $9$}%
\psfrag{0731}[lc][lc]{\matlabtextBJ $9$}%
\psfrag{0732}[lc][lc]{\matlabtextBJ $5$}%
\psfrag{0733}[lc][lc]{\matlabtextBJ $0$}%
\psfrag{0734}[lc][lc]{\matlabtextBJ $5$}%
\psfrag{0735}[lc][lc]{\matlabtextBJ $9$}%
\psfrag{0740}[rc][rc]{\matlabtextBJ $-9$}%
\psfrag{0741}[rc][rc]{\matlabtextBJ $-5$}%
\psfrag{0742}[rc][rc]{\matlabtextBJ $0$}%
\psfrag{0743}[rc][rc]{\matlabtextBJ $5$}%
\psfrag{0744}[rc][rc]{\matlabtextBJ $9$}%
\psfrag{0745}[rc][rc]{\matlabtextBJ $-9$}%
\psfrag{0746}[rc][rc]{\matlabtextBJ $-5$}%
\psfrag{0747}[rc][rc]{\matlabtextBJ $0$}%
\psfrag{0748}[rc][rc]{\matlabtextBJ $5$}%
\psfrag{0749}[rc][rc]{\matlabtextBJ $9$}%
\psfrag{0750}[rc][rc]{\matlabtextBJ $-9$}%
\psfrag{0751}[rc][rc]{\matlabtextBJ $-5$}%
\psfrag{0752}[rc][rc]{\matlabtextBJ $0$}%
\psfrag{0753}[rc][rc]{\matlabtextBJ $5$}%
\psfrag{0754}[rc][rc]{\matlabtextBJ $9$}%
\psfrag{0755}[rc][rc]{\matlabtextBJ $-9$}%
\psfrag{0756}[rc][rc]{\matlabtextBJ $-5$}%
\psfrag{0757}[rc][rc]{\matlabtextBJ $0$}%
\psfrag{0758}[rc][rc]{\matlabtextBJ $5$}%
\psfrag{0759}[rc][rc]{\matlabtextBJ $9$}%
\providecommand\matlabtextBK{\color[rgb]{0.000,0.000,0.000}\fontsize{8}{8}\selectfont\strut}%
\psfrag{0769}[tc][tc]{\matlabtextBK $d$}%
\def\matlabfragNegXTick{\mathord{\makebox[0pt][r]{$-$}}}
\providecommand\matlabtextBL{\color[rgb]{0.000,0.000,0.000}\fontsize{5}{5}\selectfont\strut}%
\psfrag{0765}[ct][ct]{\matlabtextBL $10^{0}$}%
\psfrag{0766}[ct][ct]{\matlabtextBL $10^{1}$}%
\psfrag{0767}[rc][rc]{\matlabtextBL $10^{0}$}%
\psfrag{0768}[rc][rc]{\matlabtextBL $10^{1}$}%
\journal{Physica A}

\providecommand{\norm}[1]{\vert#1\vert}

\providecommand{\mean}[1]{\langle#1\rangle}
\providecommand{\au}{A\$}
\providecommand{\us}{\$}

\begin{document}

\begin{frontmatter}

\title{Loan and nonloan flows in the Australian interbank network}

\author[unimelb]{Andrey Sokolov}
\author[unimelb]{Rachel Webster}
\author[unimelb]{Andrew Melatos}
\author[unimelb,portland]{Tien Kieu}

\address[unimelb]{School of Physics, University of Melbourne, Parkville, VIC 3010, Australia}
\address[portland]{Centre for Atom Optics and Ultrafast Spectroscopy, Swinburne University of Technology, Hawthorn, VIC 3122, Australia}

\begin{abstract}
High-value transactions between banks in Australia are settled in the Reserve Bank Information and Transfer System (RITS) administered by the Reserve Bank of Australia.
RITS operates on a real-time gross settlement (RTGS) basis and settles payments and transfers sourced from the
SWIFT payment delivery system, the Austraclear securities settlement system, and the interbank transactions entered directly into RITS.
In this paper, we analyse a dataset received from the Reserve Bank of Australia that includes all interbank transactions settled in RITS on an RTGS basis during 
five consecutive weekdays from 19 February 2007 inclusive, a week of relatively quiescent market conditions.
The source, destination, and value of each transaction are known, which allows us to separate overnight loans from other transactions (nonloans)
and reconstruct monetary flows between banks for every day in our sample.
We conduct a novel analysis of the flow stability and examine the connection between loan and nonloan flows.
Our aim is to understand the underlying causal mechanism connecting loan and nonloan flows.
We find that the imbalances in the banks' exchange settlement funds resulting from the daily flows of nonloan transactions
are almost exactly counterbalanced by the flows of overnight loans.
The correlation coefficient between loan and nonloan imbalances is about \( -0.9 \) on most days.
Some flows that persist over two consecutive days can be highly variable, but overall the flows are moderately stable in value.
The nonloan network is characterised by a large fraction of persistent flows,
whereas only half of the flows persist over any two consecutive days in the loan network.
Moreover, we observe an unusual degree of coherence between persistent loan flow values on Tuesday and Wednesday.
We probe static topological properties of the Australian interbank network and find them consistent with those observed in other countries.
\end{abstract}

\begin{keyword}
Australian interbank networks \sep
Transactional flows \sep
Overnight loans 
\end{keyword}

\end{frontmatter}

\section{Introduction}
\label{section.intro}
Financial systems are characterised by a complex and dynamic network of relationships between multiple agents.
Network analysis offers a powerful way to describe and understand the structure and evolution of these relationships;
background information can be found in \cite{kolaczyk2009statistical}, \cite{jackson2008social}, and  \cite{caldarelli2007scale}.
The network structure plays an important role in determining system stability in response to the spread of contagion,
such as epidemics in populations or liquidity stress in financial systems.
The importance of network studies in assessing stability and systemic risk has been emphasised in  \cite{schweitzer2009economic} in the context of 
integrating economic theory and complex systems research.
Liquidity stress is of special interest in banking networks.
The topology of a banking network is recognised as one of the key factors in
system stability against external shocks and systemic risks \cite{haldane2011}.
In this respect, financial networks resemble ecological networks.
Ecological networks demonstrate robustness against shocks by virtue of their continued survival
and their network properties are thought to make them more resilient against disturbances \cite{may2008ecology}.
Often they are disassortative in the sense that highly connected nodes tend to have most of their connections with weakly connected nodes (see \cite{newman2003mixing} for details).
Disassortativity and other network properties are often used to judge stability of financial networks.

There has been an explosion in empirical interbank network studies in the last years thanks largely to the introduction
of electronic settlement systems.
One of the first, reported in \cite{boss2004}, examines the Austrian interbank market, which involves about 900 participating banks.
The data are drawn from the Austrian bank balance sheet database (MAUS) and the major loan register (GKE) containing all high-value interbank loans above \euro\( 0.36\times10^6 \);
smaller loans are estimated by means of local entropy maximisation.
The authors construct a network representation of interbank payments for ten quarterly periods from 1999 to 2003.
They find that the network exhibits small-world properties and is characterised by a power-law distribution of degrees.
Specifically, the degree distribution is approximated by a power law  with the exponent \( -2.01 \) for degrees \( \gtrsim40 \).
This result, albeit with different exponents, holds for the in- and out-degree distributions too (the exponent is \( -3.1 \) for out-degrees and \( -1.7 \) for in-degrees).
A recent study of transactional data from the Austrian real-time interbank settlement system (ARTIS) reported in \cite{kyriakopoulos2009network} 
demonstrates a strong dependence of network topology on the time-scales of observation,
with power-law tails exhibiting steeper slopes when long time-scales are considered.

The network structure of transactions between Japanese banks, logged by the Bank of Japan Financial Network system (BOJ-NET), is analysed in \cite{inaoka2004fractal}.
The authors consider several monthly intervals of data from June 2001 to December 2002 and construct monthly networks of interbank links corresponding to 21 transactions or more, 
i.e.\ one or more transaction per business day on average.
Truncating in this way eliminates about 200 out of 546 banks from the network.
The resulting monthly networks have a low connectivity of 3\% and a scale-free cumulative distribution of degrees with the exponent \( -1.1 \).

More than half a million overnight loans from the Italian electronic broker market for interbank deposits (e-MID),
covering the period from 1999 to 2002, are analysed in \cite{de2006fitness}.
There are about 140 banks in the network, connected by about 200 links.
The degree distribution is found to exhibit fat tails with power-law exponent \( 2.3 \) (\( 2.7 \) for in-degrees and \( 2.15 \) for out-degrees),
the network is disassortative, with smaller banks staying on its periphery. 
In a related paper \cite{iori2007trading}, the authors make use of the same dataset to uncover liquidity management
strategies of the participating banks, given the reserve requirement of 2\% on the 23rd of each month imposed by the central bank.
Signed trading volumes are used as a proxy for the liquidity strategies and their correlations are analysed.
Two distinct communities supporting the dichotomy in strategy are identified by plotting the correlation matrix as a graph.
The two communities are mainly composed of large and small banks respectively.
On average, small banks serve as lenders and large banks as borrowers, but the strategies reversed in July 2001,
when target interest rates in the Euro area stopped rising and started to decrease.
The authors also note that some mostly small banks tend to maintain their reserves through the maintenance period.
The evolution of the network structure over the monthly maintenance period is examined in \cite{iori2008network}.

A study of the topology of the Fedwire network, a real-time gross settlement (RTGS) system 
operated by the Federal Reserve System in the USA, is reported in \cite{soramaki2007}.
The study covers 62 days in the 1st quarter of 2004, during which time Fedwire comprised more than 7500 participants and settled \( 3.45\times10^5 \) payments daily with total value {\us}1.3 trillion.
It reveals that Fedwire is a small-world network with low connectivity (0.3\%), moderate reciprocity (22\%), and a densely connected sub-network of 25 banks
responsible for the majority of payments. 
Both in- and out-degree distributions follow a power law for degrees $\gtrsim10$  (exponent \( 2.15 \) for in-degrees and \( 2.11 \) for out-degrees).
The network is disassortative, with the correlation of out-degrees equal to \( -0.31 \).
The topology of overnight loans in the federal funds market in the USA is examined in \cite{bech2010},
using a large dataset spanning 2415 days from 1999 to 2006.
It is revealed that the overnight loans form a small-world network, which is sparse (connectivity 0.7\%), disassortative (assortativity ranging from \( -0.06 \) to \( -0.28 \)),
and has low reciprocity of 6\%.
The reciprocity changes slowly with time and appears to follow the target interest rate over the period of several years.
A power law is the best fit for the in-degree distribution, but the fit is only good for a limited range of degrees.
A  negative binomial distribution, which requires two parameters rather than one for a power law, fits the out-degree distribution best.

A comprehensive survey of studies of interbank networks is given in \cite{imakubo2010transaction}.
The number of interbank markets being analysed continues to increase.
For example, a study of the interbank exposures in Brazil for the period from 2004 to 2006 was reported in \cite{cajueiro2008role}.
A topological analysis of money market flows logged in the Danish large-value payment system (Kronos) in 2006
was reported in \cite{rordam2008topology}, 
where customer-driven transactions are compared with the bank-driven ones.
Empirical network studies have been used to guide the development of a network model of the interbank market based on the interbank credit
lending relationships \cite{li2010}.

Establishing basic topological features of interbank networks is essential for understanding these complex systems.
Fundamentally, however, interbank money markets are flow networks, in which links between the nodes   correspond to monetary flows.
The dynamics of such flows has not been examined in depth in previous studies,
which mostly viewed interbank networks as static or slowly varying.
But the underlying flows are highly dynamic and complex.
Moreover, monetary flows are inhomogeneous; loan flows are fundamentally different from the flows of other payments.
Payments by the banks' customers and the banks themselves cause imbalances in the exchange settlement accounts of the banks.
For some banks, the incoming flows exceed the outgoing flows on any given day; for other banks, the reverse is true.
Banks with excess reserves lend them in the overnight money market to banks with depleted reserves.
This creates interesting dynamics: payment flows cause imbalances, which in turn drive compensating flows of loans.
Understanding this dynamic relationship is needed for advancing  our ability to model interbank markets effectively.

In this paper, our objective is to define empirically the dynamics of interbank monetary flows.
Unlike most studies cited above, we aim to uncover the fundamental causal relationship between the flows of overnight loans and other payments. 
We choose to specialise in the Australian interbank market, where we have privileged access to a high-quality dataset provided by the Reserve Bank of Australia (RBA).
Our dataset consists of transactions settled in the period from 19 to 23 February 2007 in the Australian interbank market.
We separate overnight loans and other payments (which we call nonloans) using a standard matching procedure.
The loan and nonloan transactions settled on a given day form the flow networks, which are the main target of our statistical analysis.
We compare the topology and variation of the loan and nonloan networks and reveal the causal mechanism that ties them together.
We investigate the dynamical stability of the system by testing how individual flows vary from day to day.
Basic network properties such as the degree distribution and assortativity are examined as well.

\section{Data}
\label{section.data}
High-value transactions between Australian banks are settled via the Reserve Bank Information and Transfer System (RITS)
operated by the RBA since 1998 on an RTGS basis \citep{gallagher2010}.
The transactions are settled continuously throughout the day by crediting and debiting the exchange settlement accounts
held by the RBA on behalf of the participating banks. 
The banks' exchange settlement accounts at the RBA are continuously monitored to ensure liquidity,
with provisions for intra-day borrowing via the intra-day liquidity facility provided to the qualifying banks by the RBA.
This obviates the need for a monthly reserve cycle of the sort maintained by Italian banks as discussed in \cite{iori2008network}.
The RITS is used as a feeder system for transactions originating from SWIFT\footnote{Society for Worldwide Interbank Financial Telecommunication} 
and Austraclear for executing foreign exchange
and securities transactions respectively.
The member banks can also enter transactions directly into RITS.
The switch to real-time settlement in 1998 was an important reform which protects the payment system
against systemic risk, since transactions can only be settled if the paying banks possess sufficient funds in their exchange settlement accounts.
At present, about \( 3.2\times10^4 \) transactions are settled per day, with total value around {\au}168 billion.

The data comprise all interbank transfers processed on an RTGS basis by the RBA during the week of 19 February 2007.
During this period, 55 banks participated in the RITS including the RBA.
The dataset includes transfers between the banks and the RBA, such as RBA's intra-day repurchase agreements and money market operations.
The real bank names are obfuscated (replaced with labels from A to BP) for privacy reasons, 
but the obfuscated labels are consistent over the week.
The transactions are grouped into separate days, but the time stamp of each transaction is removed. 

\begin{table}
\centering
\begin{tabular}{ccc}
\hline
Date   & Volume & Value ({\au}\( 10^9 \))\\
\hline
19-02-2007 & 19425    & \phantom{0}82.2506 \\
20-02-2007 &  27164   & 206.1023 \\
21-02-2007 &  24436   & 161.9733  \\
22-02-2007 &  25721   & 212.1350 \\
23-02-2007 &  26332   & 184.9202 \\
\hline
\end{tabular}
\caption{The number of transactions (volume) and their total value (in units of {\au}\( 10^9 \)) for each day.}
\label{total values}
\end{table}

\begin{figure}[t]
\includegraphics{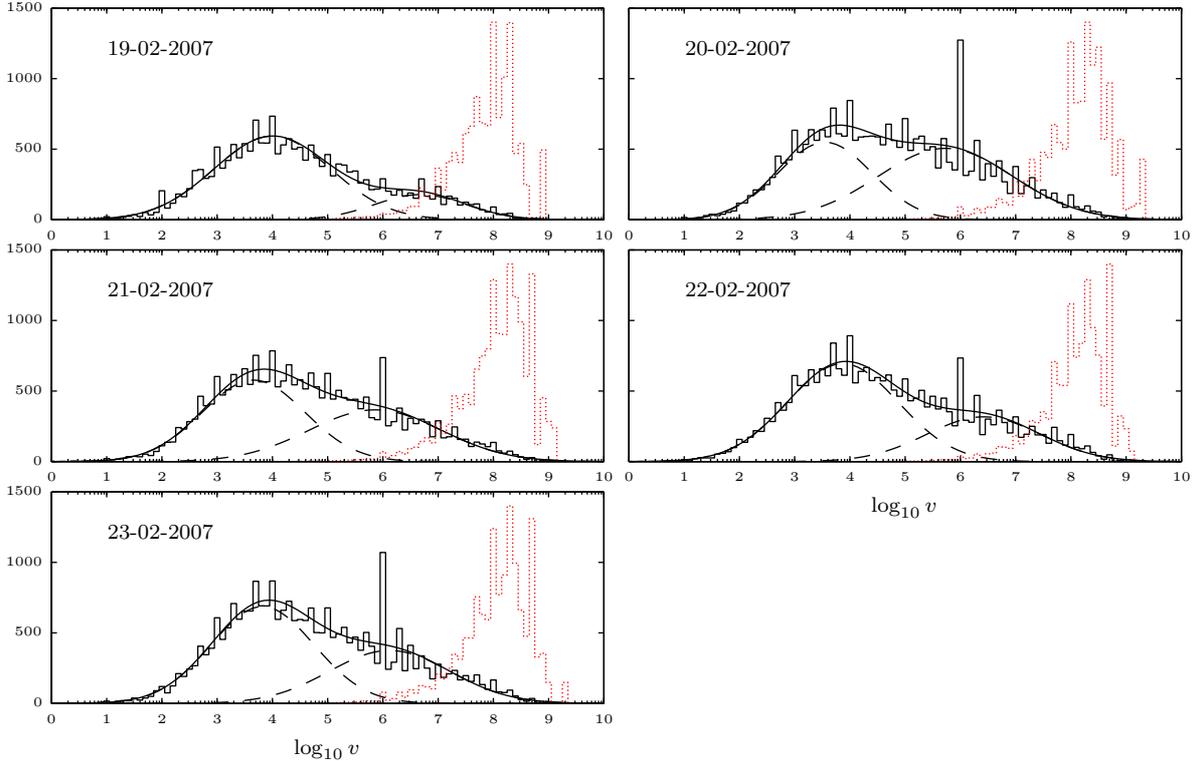}
\caption{The distribution of transaction values \( v \) (in Australian dollars) 
on a logarithmic scale, with bin size \( \Delta\log_{10}v=0.1 \);
the vertical axis is the number of transactions per bin.
Components of the Gaussian mixture model are indicated by the dashed curves; the solid curve is the sum of the two components.
The dotted histogram shows the relative contribution of transactions at different values to the total value
(to compute the dotted histogram we multiply the number of transactions in a bin by their value).
}
\label{value histograms}
\end{figure}

During the week in question, around \( 2.5\times10^4 \) transactions were settled per day, with the total
value of all transactions rising above {\au}\( 2\times10^{11} \) on Tuesday and Thursday.
The number of transactions (volume\footnote{The term ``volume'' is sometimes used to refer to the combined dollar amount of transactions. 
In this paper, we only use the term ``volume'' to refer to the number of transactions and ``total value'' to refer to the combined dollar amount.
This usage follows the one adopted by the RBA \cite{gallagher2010}.
}) and the total value 
(the combined dollar amount of all transactions) for each day are given in Table~\ref{total values}.
Figure~\ref{value histograms} shows the distribution of transaction values on a logarithmic scale.
Local peaks in the distribution correspond to round values. The most pronounced peak occurs at {\au}\( 10^6 \).

In terms of the number of transactions, the distribution consists of two approximately log-normal
components, with lower-value transactions being slightly more numerous.
The standard entropy maximisation algorithm for a Gaussian mixture model with two components \citep{mclachlan2000finite} produces a satisfactory fit
with the parameters indicated in Table~\ref{GMM}.
The lower- and higher-value components are typically centred around {\au}\( 10^4 \) and {\au}\( 10^6 \) respectively.
The high-value component is small on Monday (19-02-2007)
but increases noticeably on subsequent days, while the low-value component diminishes.
By value, however, the distribution is clearly dominated by transactions above {\au}\( 10^6 \), 
with the highest contribution from around {\au}\( 2\times10^8 \).

\begin{table}
\centering
\begin{tabular}{ccccccc}
\hline
Date   & \multicolumn{3}{c}{Component 1} & \multicolumn{3}{c}{Component 2} \\
 & \( \mean{u} \) & \( \sigma_u^2 \) & \( P \) & \( \mean{u} \) & \( \sigma_u^2 \) & \( P \) \\
\hline
19-02-2007 & 4.00 & 1.12 & 0.81 & 6.68 & 0.68 & 0.19 \\
20-02-2007 & 3.55 & 0.72 & 0.43 & 5.73 & 1.49 & 0.57 \\
21-02-2007 & 3.66 & 0.86 & 0.55 & 5.86 & 1.43 & 0.45 \\
22-02-2007 & 3.87 & 1.01 & 0.68 & 6.42 & 1.07 & 0.32 \\ 
23-02-2007 & 3.82 & 0.87 & 0.61 & 6.12 & 1.19 & 0.39 \\ 
\hline
\end{tabular}
\caption{Mean \( \mean{u} \), variance \( \sigma_u^2 \), and mixing proportion \( P \) of the Gaussian mixture components
shown in Figure~\ref{value histograms}
(\( u=\log_{10}v \), where \( v \) is value).}
\label{GMM}
\end{table}

\section{Overnight loans}
The target interest rate of the RBA during the week of our sample was \( r_t=6.25\% \) per annum. 
If the target rate is known, it is easy to extract the overnight loans from the data by identifying reversing transactions on consecutive days.
A hypothetical interest rate can be computed for each reversing transaction and compared with the target rate.
For instance, suppose a transaction of value \( v_1 \) from bank A to bank B on day 1 
 reverses with value \( v_2 \), from bank B to bank A, on day 2.
These transactions are candidates for the first and second legs of an overnight loan from A to B.
The hypothetical interest rate for this pair of transactions is given by \( r_h = 100\%\times365\times(v_2-v_1)/v_1 \);
note that the quoted target rate is per annum.
Since large banks participate in many reversing transactions that can qualify as loans,
we consider all possible hypothetical pairs and prefer the one that gives \( r_h \) closest to the target rate.
The algorithm for loan extraction is applied from Monday to Thursday;
loans issued on Friday cannot be processed since the next day is not available.
A similar procedure was pioneered by Furfine \cite{furfine2003interbank}; see also \cite{ashcraft2007systemic}.

\begin{figure}[t]
\begin{center}
\includegraphics{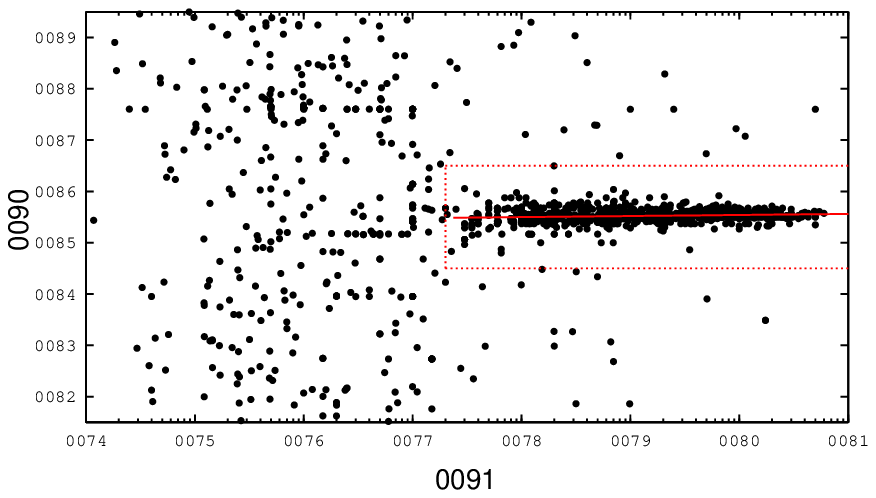}
\caption{Hypothetical interest rate \( r_h \) versus value of the first leg of the transaction pairs detected by our algorithm,
with no restrictions on value or interest rate.
The dotted rectangle contains the transactions that we identify as overnight loans.
The least-squares fit is shown with a solid red line.
}\label{loans scatter}
\end{center}
\end{figure}

The application of the above algorithm results in the scatter diagram shown in Figure~\ref{loans scatter}.
There is a clearly visible concentration of the reversing transaction pairs in the region  \( v>2\times10^5 \) and \( |r_t-r_h|<0.5\% \) (red box).  
We identify these pairs as overnight loans.
Contamination from nonloan transaction pairs that accidentally give a hypothetical rate close to the target rate
is insignificant. 
By examining the adjacent regions of the diagram, i.e.\ \( v>2\times10^5 \) and \( r_h \) outside of the red box, we estimate the contamination to be less than 2\%
(corresponding to \( \le5 \) erroneous identifications per day).
It is also possible that some genuine loans fall outside our selection criteria. 
However, it is unlikely that overnight interest rates are very different from the target rate;
and the lower-value transactions (below {\au}\( 10^4 \)), even if they are real loans, contribute negligibly to the total value.

\begin{table}
\centering
\begin{tabular}{ccccc}
\hline
Date  & Volume & Value ({\au}\( 10^9 \)) & Loan fraction\\
\hline
19-02-2007 &185   & \phantom{0}7.50 & 9.12\% \\
20-02-2007  & 221  & \phantom{0}9.18 & 4.45\% \\
21-02-2007 &  226  & 11.08 & 6.84\% \\
22-02-2007 & 265  & 14.93 & 7.04\% \\
\hline
\end{tabular}
\caption{Statistics of the overnight loans identified by our algorithm:
the number of loans (volume), the total value of the first leg of the loans (in units of {\au}\( 10^9 \)), and
the fraction of the total value of the loans (first legs only) with respect to the total value of all transactions on a given date.}
\label{loans}
\end{table}

\begin{figure}[t!]
{\centering
\subfloat[]{\includegraphics{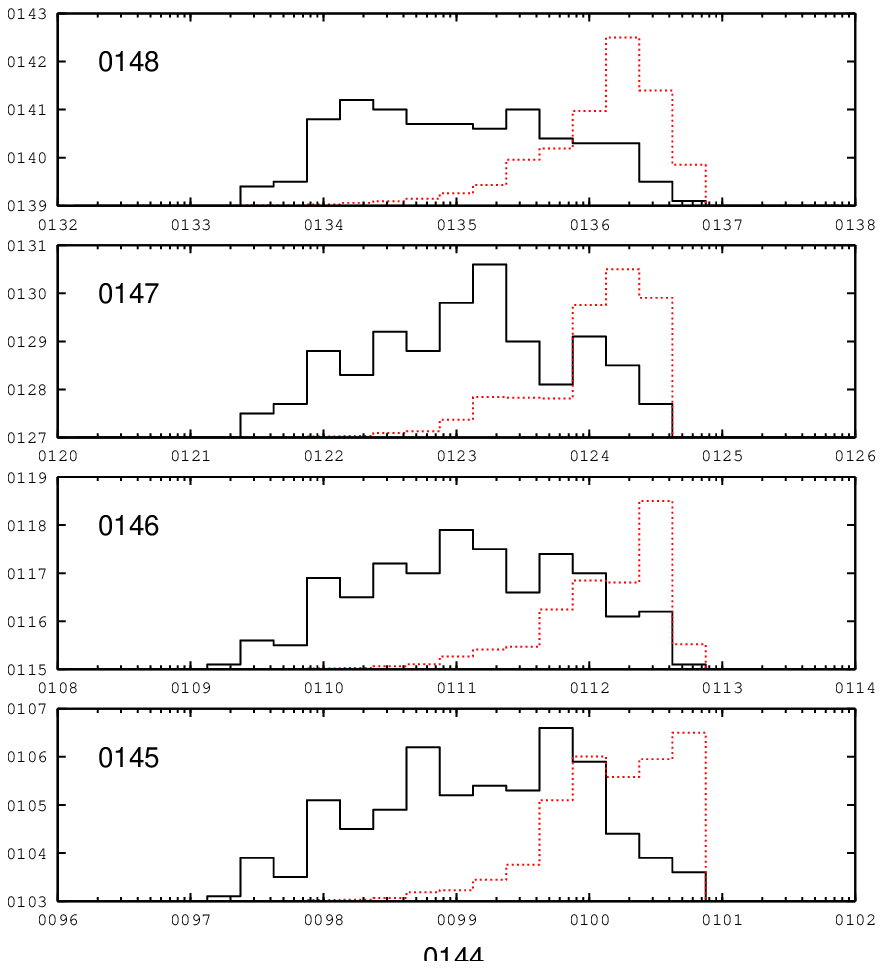}\label{loans values}}
\subfloat[]{\includegraphics{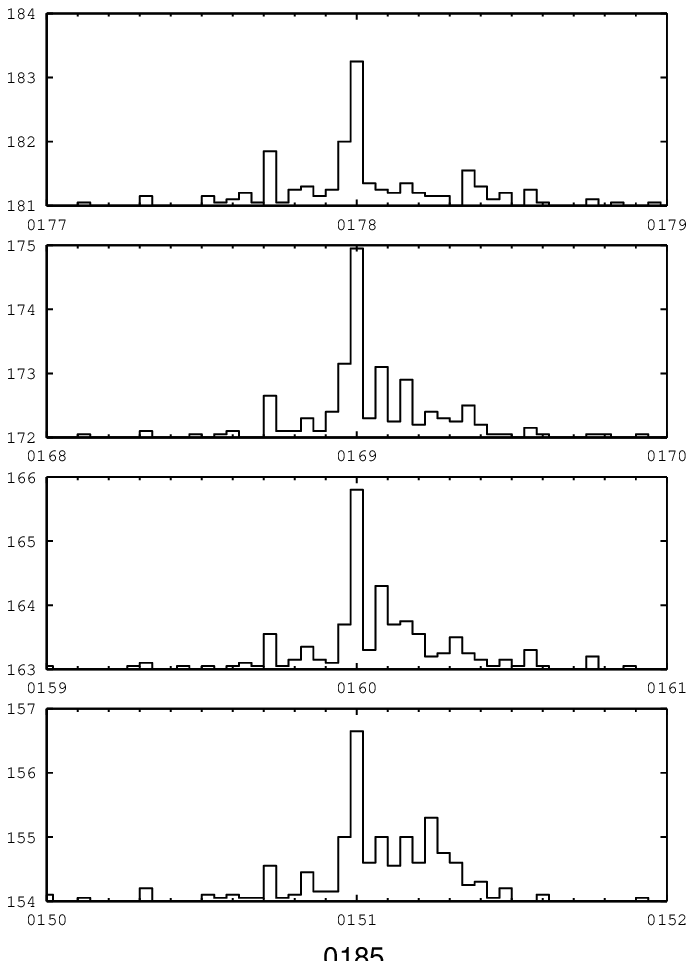}\label{loans rates}}
}
\caption{\subref{loans values} The distribution of loan values \( v \) on a logarithmic scale. 
The vertical axis is the number of loans per bin for bin size \( \Delta\log_{10}v=0.25 \).
The dotted line is the same distribution multiplied by the value corresponding to each bin (in arbitrary units).
The date of the first leg of the loans is indicated.
\subref{loans rates} The distribution of loan interest rates \( r_h \). 
The vertical axis is the number of loans per bin for bin size \( \Delta r_h=0.01 \).
The date of the first leg of the loans is indicated. 
The mean and standard deviation are 6.25\% and 0.08\%
on Monday (19-02-2007),
and 6.26\% and 0.07\% on the other days.}
\end{figure}

We identify 897 overnight loans over the four days.
A daily breakdown is given in Table~\ref{loans}.
Here and below, we refer to the first leg of the overnight loans as simply loans and to all other transactions as nonloans.
The loans constitute less than 1\% of all transactions by number and up to 9\% by value (cf. Tables~\ref{total values} and~\ref{loans}). 
The distribution of loan values and interest rates is shown in Figures~\ref{loans values} and~\ref{loans rates}.
The interest rate distribution peaks at the target rate 6.25\%.
The mean rate is within one basis point (0.01\%) of the target rate, while the standard deviation is about 0.07\%.
The average interest rate increases slightly with increasing value of the loan;
a least-squares fit yields  \( r_h=6.248+0.010\log_{10}(v/\mathrm{\au}10^6) \).

The same technique can be used to extract two-day and longer-term loans 
(up to four-day loans for our sample of five consecutive days).
Using the same selection criteria as for the overnight loans, our algorithm detects
27, 67, and 24 two-day loans,
with total values  {\au}1.3, {\au}2.2, and {\au}1.4 billion, on Monday, Tuesday, and Wednesday, respectively.
The total value of the two-day loans 
is 1.5\%,  1.0\%, and 0.9\% of the total transaction values  on these  days respectively.

\section{Nonloans}
\begin{figure*}[t]
\includegraphics{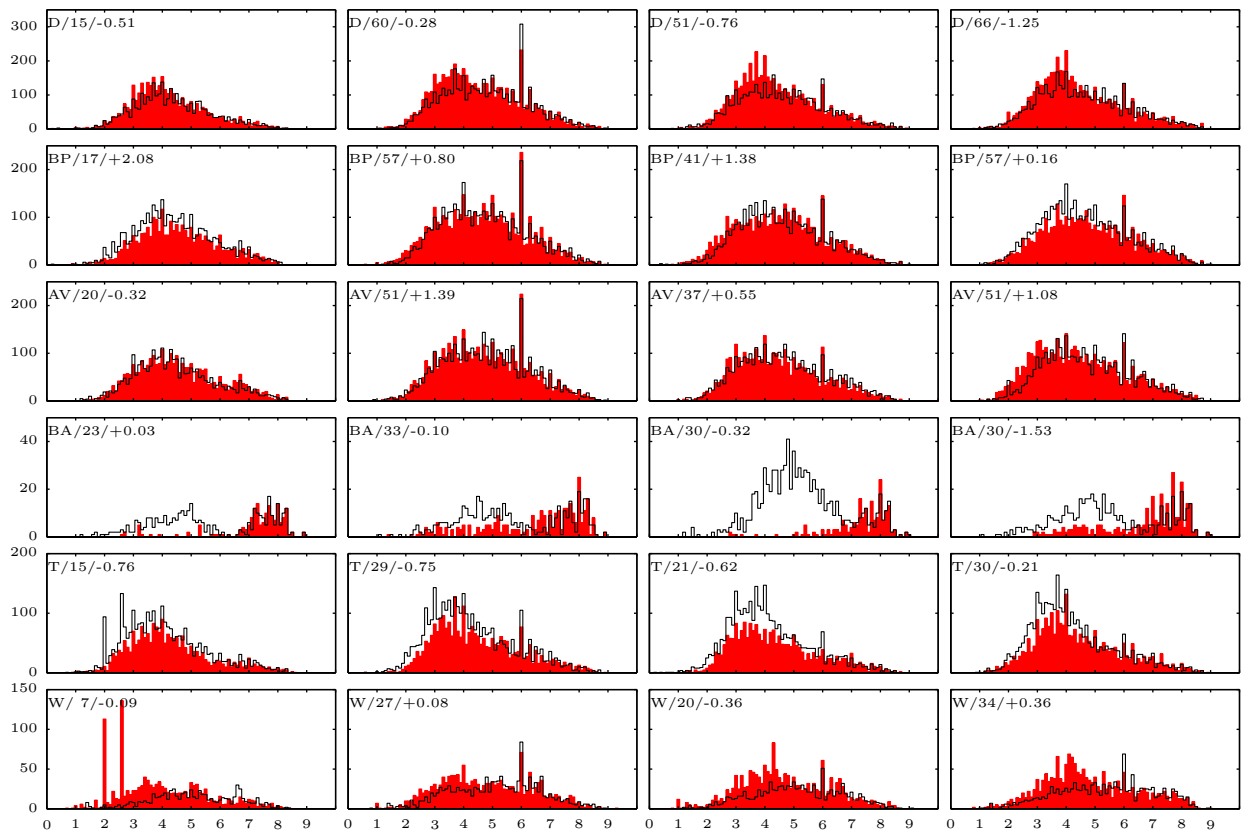}
\caption{The distribution of nonloan transaction values of the six largest banks for Monday through Thursday (from left to right);
the banks are selected by the combined value of incoming and outgoing transactions over the entire week.
Black and red histograms correspond to 
incoming (bank is the destination) and  outgoing (bank is the source) transactions; red histograms are filled in to improve visibility.
The banks' anonymous labels, the combined daily value of the incoming and outgoing transactions, and the daily imbalance (incoming minus outgoing) are quoted 
at the top left of each panel (in units of {\au}\( 10^9 \)). The horizontal axis is the logarithm of value in {\au}.}
\label{individual distributions}
\end{figure*}

We display the distributions of the  
incoming and outgoing nonloan transactions,
for which the bank is the destination and the source respectively,
for the six largest banks in Figure~\ref{individual distributions}.
The distributions are similar to the total distribution shown in Figure~\ref{value histograms},
 with the notable exception of BA (see below).
There is also an unusually large number of {\au}106 and {\au}400  transactions from W to T on Monday. 
Note that the daily imbalance for each bank is mostly determined by the highest value transactions;
large discrepancies between incoming and outgoing transactions at lower values are less relevant.

The distribution for BA is clearly bimodal; it contains an unusually high proportion of transactions greater than {\au}\( 10^6 \).
Moreover, below {\au}\( 10^6 \), incoming transactions typically outnumber outgoing ones by a large amount.
BA is also involved in many high value transactions that reverse on the same day. 
These transactions probably correspond to the central bank's repurchase agreements, which facilitate intra-day liquidity of the banks \citep{rbarepos}.

The banks shown in Figure~\ref{individual distributions} are also the largest in term of the number of transactions,
with the exception of BA.
The rank order by the number of transactions matches that by value.
For D, which is the largest, the number of nonloan transactions reaches 48043 over the week.
By the number of transactions, the order of the top twelve banks is D, BP, AV, T, W, AH, AF, U, AP, BI, BA, P.
By value, the order is  D, BP, AV, BA, T, W, BG, U, A, AH, AB, BM.
The situation is similar when considering the overnight loans. 
By value, AV, D, BP, and T dominate. 
For these four banks, weekly total loans range from {\au}11.5 to {\au}18 billion
and number from 254 to 399.
For the other banks
the total loan value is less than {\au}3 billion. 

In view of the discussion above, it is noteworthy that Australia's retail banking system is dominated by four big banks 
(ANZ, CBA, NAB, and WBC)\footnote{Australia and New Zealand Banking Group, 
Commonwealth Bank of Australia, National Australia Bank, and Westpac Banking Corporation.}
that in February 2007 accounted for 65\% of total resident assets,
according to statistics published by Australian Prudential Regulation Authority (APRA); see http://www.apra.gov.au for details.
The resident assets of the big four exceeded {\au}225 billion each,
well above the next largest retail bank, St George Bank Limited\footnote{%
In December 2008, St George Bank became a subsidiary of Westpac  Banking Corporation.}
 ({\au}93 billion).
The distinction between the big four and the rest of the banks in terms of cash and liquid assets at the time was less clear,
with Macquarie Bank Limited in third position with {\au}8 billion.
According to APRA, cash and liquid assets of the big four and Macquarie Bank Limited
accounted for 56\% of the total.

\section{Loan and nonloan imbalances}

In order to maintain liquidity in their exchange settlement accounts,
banks ensure that incoming and outgoing transactions roughly balance.
However, they do not control most routine transfers, 
which are initiated by account holders.
Therefore, the imbalances arise.
On any given day, the nonloan imbalance of bank \( i \) is given by
\begin{equation}
\Delta v_i
=
-\sum_j\sum_k
v_k(i,j)
+\sum_j\sum_k
v_k(j,i),
\end{equation}
where \( \{v_k(i,j)\}_k \) is a list of values of individual nonloan transaction from bank \( i \) to bank \( j \), settled on the day.
The nonloan imbalances are subsequently compensated by overnight loans traded on the interbank money market.
The loan imbalances are defined in the same way using transactions corresponding to the first leg of the overnight loans.
Note that we do not distinguish between the loans initiated by the banks themselves
and those initiated by various institutional and corporate customers.
For instance, if the funds of a corporate customer are depleted, this customer may borrow overnight
 to replenish the funds.
In this case, the overnight loan is initiated by an account holder, who generally has no knowledge of
the bank's net position.
Nevertheless, the actions of this account holder in acquiring a loan reduce the bank's imbalance,
provided that the customer deposits the loan in an account with the same bank.

\begin{figure}
\centering
\includegraphics{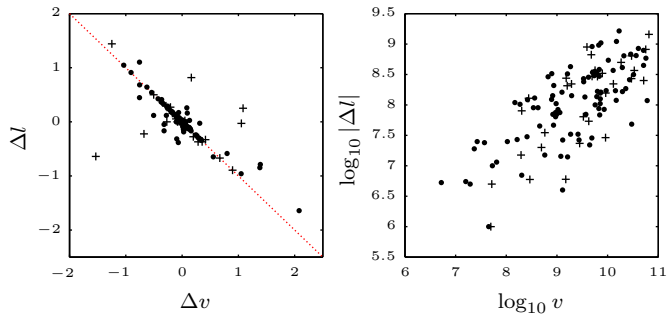}
\caption{Left: loan imbalance \( \Delta l \) vs nonloan imbalance \( \Delta v \) for individual banks and days of the week (in units of {\au}\( 10^9 \)).
Right: the absolute value of loan imbalance \( |\Delta l| \) vs nonloan total value (incoming plus outgoing transactions)
for individual banks and days of the week. Thursday data are marked with crosses.
}
\label{imbalances correlation}
\end{figure}

\begin{table*}
\centering
\begin{tabular}{lcccccccc}
\hline
 & \multicolumn{2}{c}{19-02-2007} & \multicolumn{2}{c}{20-02-2007} & \multicolumn{2}{c}{21-02-2007} & \multicolumn{2}{c}{22-02-2007} \\
 & nonloans & loans & nonloans & loans & nonloans & loans & nonloans & loans \\
\hline
D  & \( -0.51 \) & \( +0.12 \) & \( -0.28 \) & \( +0.12 \) &  \( -0.76 \) & \( +0.45 \) &  \( -1.25 \) & \( +1.44 \) \\
BP & \( +2.08 \) & \( -1.64 \) & \( +0.80 \) & \( -0.59 \) &  \( +1.38 \) & \( -0.85 \) &  \( +0.16 \) & \( +0.82 \) \\
AV & \( -0.32 \) & \( -0.17 \) & \( +1.39 \) & \( -0.79 \) &  \( +0.55 \) & \( -0.65 \) &  \( +1.08 \) & \( +0.25 \) \\
BA & \( +0.03 \) & \( -0.19 \) & \( -0.10 \) & \( -0.31 \) &  \( -0.32 \) & \( -0.05 \) &  \( -1.53 \) & \( -0.64 \) \\
T  & \( -0.76 \) & \( +1.10 \) & \( -0.75 \) & \( +0.68 \) &  \( -0.62 \) & \( +0.64 \) &  \( -0.21 \) & \( +0.27 \) \\
W  & \( -0.09 \) & \( +0.07 \) & \( +0.08 \) & \( +0.26 \) &  \( -0.36 \) & \( +0.41 \) &  \( +0.36 \) & \( -0.37 \) \\
\hline
\end{tabular}
\caption{Loan and nonloan imbalances for the six largest banks (in units of {\au}\( 10^9 \)).}
\label{imbalances and loans}
\end{table*}

The loan and nonloan imbalances for the six largest banks are given in Table~\ref{imbalances and loans}.
The data generally comply with our assumption that the overnight loans compensate the daily imbalances of the nonloan transactions.
The most obvious exception is for BA on Thursday (22-02-2007), where a large negative nonloan imbalance is accompanied by a
sizable loan imbalance that is also negative.
Taking all the banks together, there is a strong anti-correlation between loan and nonloan imbalances on most days.
We see this clearly in Figure~\ref{imbalances correlation}.
The Pearson correlation coefficients for Monday through Thursday are \( -0.93 \),   \( -0.88 \),   \( -0.95 \),   \( -0.36 \). 
It is striking to observe that many points fall close to the perfect anti-correlation line.
The anti-correlation is weaker on Thursday (crosses in Figure~\ref{imbalances correlation}),
mostly due to BA and AV.

A correlation also exists between the absolute values of loan imbalances and the nonloan total values (incoming plus outgoing nonloan transactions);
the Pearson  coefficients are \( 0.74 \), \( 0.75 \), \( 0.66 \), \( 0.77 \) for Monday through Thursday.
This confirms the intuitive expectation that  larger banks tolerate larger loan imbalances.

\section{Flow variability}
\label{flow variability section}
For each individual source and destination, we define the nonloan flow as
the totality of all nonloan transactions from the given source to the given destination on any given day.
The value of the flow is the sum of the nonloan transaction values and the direction is from the source to the destination.
On any given day, the value of the flow from bank \( i \) to bank \( j \) is defined by 
\begin{equation}
v_{\textrm{flow}}(i,j)=\sum_{k} v_k(i,j),
\end{equation}
where 
\( \{v_k(i,j)\}_k \) is a list of values of individual nonloan transaction from \( i \) to \( j \) on the day.
For example, all nonloan transactions from D to AV on Monday form a nonloan flow from D to AV on that day.
The nonloan transactions in the opposite direction, from AV to D, form another flow.
A flow has zero value if the number of transactions is zero. 
Typically, for any two large banks there are two nonloan flows between them.
The loan flows are computed in a similar fashion.

\subsection{Nonloan flows}

There are 55 banks in the network, resulting in \( N_\textrm{flow}=2970 \) possible flows.
The actual number of flows is much smaller.
The typical number of nonloan flows is \( \sim800 \) on each day (the actual numbers are  804,  791, 784,  797).
Even though the number of nonloan flows does not change significantly from day to day,
we find that only about 80\% of these flows persist for two days or more.
The other 20\% are replaced by different flows, i.e.\ with a different source and/or destination, on the following day.
Structurally speaking, the network of nonloan flows changes by 20\% from day to day.
However, persistent flows carry more than 96\% of the total value.

\begin{figure*}[t]
\includegraphics{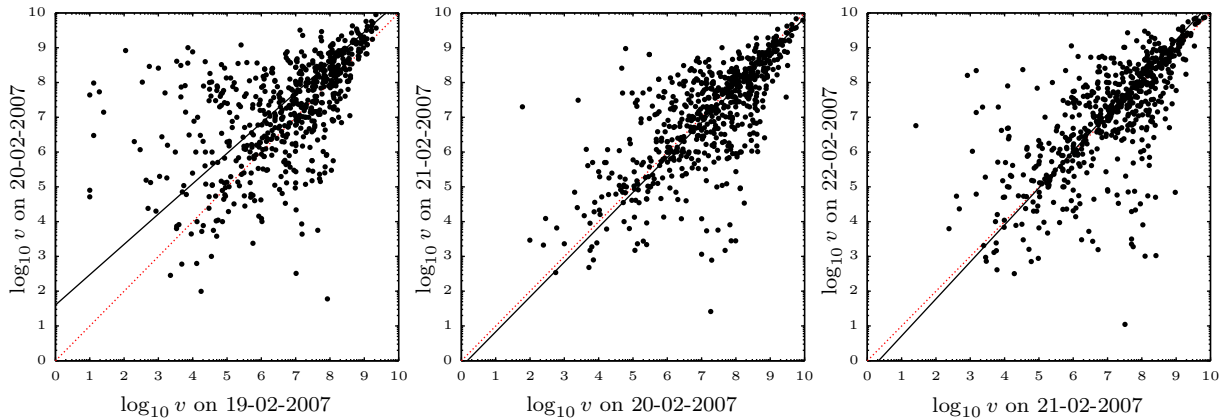}
\caption{Nonloan flow value pairs on one day (horizontal axis) and the next (vertical axis).
Only flows present on both days are considered.
Flows that do not change lie on the diagonal (red dotted line).
The solid line is the weighted orthogonal least squares fit to the scatter diagram;
the weights have been defined to emphasize points corresponding to large flows.
}
\label{flow value correlations}
\end{figure*}

Even when the flow is present on both days, its value is rarely the same.
Given that 80\% of the network is structurally stable from day to day, 
we assess variability of the network by considering
persistent flows and their values on consecutive days.
Figure~\ref{flow value correlations} shows the pairs of persistent flow values for Monday and Tuesday, Tuesday and Wednesday, and Wednesday and Thursday.
If the flow values were the same, the points in Figure~\ref{flow value correlations}
would lie on the diagonals.
We observe that the values of some flows vary significantly, especially when comparing Monday and Tuesday.
Moreover, there is a notable systematic increase in value of the flows from Monday to Tuesday by a factor of several,
which is not observed on the other days.
For each pair of days shown in Figure~\ref{flow value correlations}, we compute the Pearson correlation coefficient, which gives
0.53 for Monday and Tuesday, 0.70 for Tuesday and Wednesday, and 0.68 for Wednesday and Thursday.

To characterize the difference between the flows on different days more precisely, we compute the Euclidean distance between normalised flows on consecutive days.
We reorder the adjacency matrix \( \{v_{\textrm{flow}}(i,j)\}_{ij} \) of the flow network on day \( d \) as an \( N_\textrm{flow} \)-dimensional vector \( \mathbf{v}_d \)
representing a list of all flows on day \( d \)  (\( d=1,2,\ldots,5 \)).
For each pair of consecutive days we compute the Euclidean distance between normalized vectors \( \mathbf{v}_d/\norm{\mathbf{v}_d} \) 
and \( \mathbf{v}_{d+1}/\norm{\mathbf{v}_{d+1}} \), which gives 0.62, 0.50, 0.50 for all flows
and  0.61, 0.49, 0.49 for persistent flows (the latter are computed by setting non-persistent flows to zero on both days).
Since the flow vectors are normalized,  these quantities measure random flow discrepancies
while systematic deviation such as between the flows on Monday and Tuesday are ignored.
For two vectors of random values uniformly distributed in interval \( (0,1) \), the expected Euclidean distance is 0.71 
and the standard deviation is 0.02 for the estimated number of persistent nonloan flows of \( 640 \).
So the observed variability of the nonloan flows is smaller than what one might expect if the flow values were random.

\begin{figure*}[t]
\includegraphics{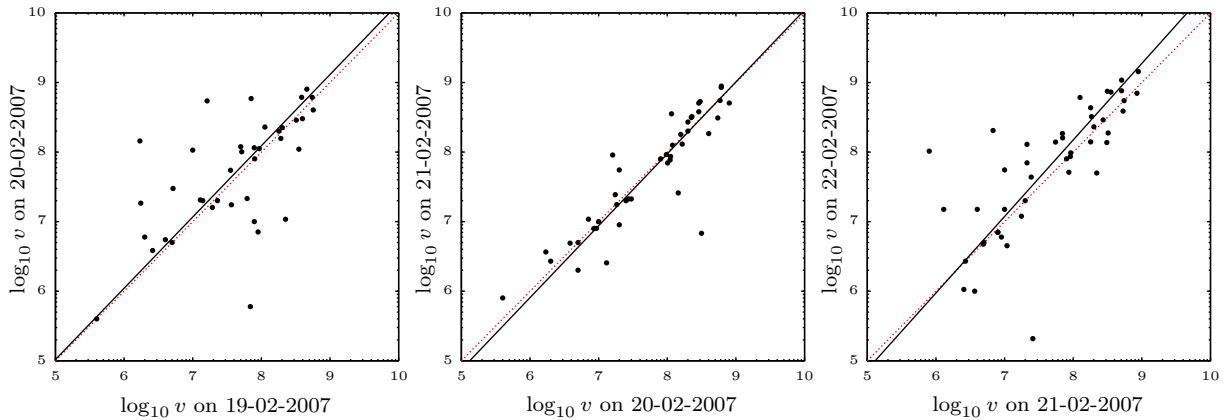}
\caption{As for Figure~\ref{flow value correlations} but for loan flows.
}
\label{loan flow value correlations}
\end{figure*}

\subsection{Loan flows}
Variability of the loan flows is equally strong.
The number of loan flows varies from 69 to 83 (actual numbers are 69, 75, 77, 83).
Only about 50\% of these flows are common for any two consecutive days.
Moreover, persistent flows carry only about 65\% of the total value of the loan flows on any given day, cf.\ 80\% of nonloan flows.
For persistent loan flows, the Pearson correlation coefficients are 0.63, 0.90, and 0.76 for the consecutive pairs of days starting with Monday and Tuesday.
The correlation is generally similar to that of the nonloan flows,
with the notable exception of the loan flows on Tuesday and Wednesday,
when the sub-network of persistent loan flows appears to be more stable.

The Euclidean distances between the normalized loan flows for each pair of consecutive days are 
0.85, 0.68, 0.73 for all flows and 0.63, 0.44, and 0.44 for persistent flows.
For two vectors of random values uniformly distributed in interval \( (0,1) \), the expected Euclidean distance is 0.7
and the standard deviation is 0.1 for the estimated number of persistent loan flows of \( 40 \).
So the observed variability of the persistent loan flows is much smaller than what one might expect if the flow values were random.

\subsection{Relation between nonloan and loan flows}
Some loan flows do not have corresponding
nonloan flows between the same nodes on the same day. 
These flows carry about 14\% of loan value on Monday, and about 7\% on Tuesday through Thursday.
Nonloan flows that have corresponding loan flows account for 35\% to 48\%
of all nonloan flows by value, even though the number of these flows is less than 10\% of the total.

\begin{figure}[t]
\centering
\includegraphics{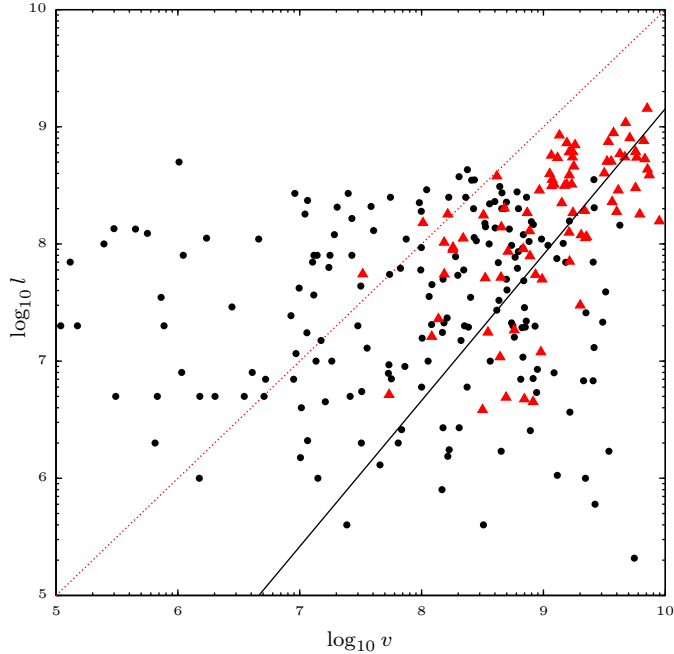}
\caption{Loan flow values versus nonloan flow values combined over four days.
Triangles correspond to loan flows with three or more transactions per flow.
The solid line is the orthogonal least squares fit to the scatter diagram;
the weighting is the same as in Figure~\ref{flow value correlations}.
}
\label{loan nonloan flow value correlations}
\end{figure}

To improve the statistics, we aggregate the flows on all four days.
Figure~\ref{loan nonloan flow value correlations} shows nonloan and corresponding loan flow values.
We fail to find any significant correlation between loan and nonloan flows (Pearson coefficient is 0.3).
The correlation improves  if we restrict the loan flows to those consisting of three transactions or more;
such flows mostly correspond to large persistent flows.
In this case the Pearson coefficient increases to 0.6;
banks that sustain large nonloan flows can also sustain large loan flows, even though 
the loan flows on average are an order of magnitude lower than the corresponding nonloan flows.
The lack of correlation when all loans are aggregated is due to the presence of many large loans that are not
accompanied by large nonloan transactions, and vice versa.

\section{Net flows}
The net flow between any two banks is defined as the difference of the opposing flows between these banks.
The value of the net flow equals the absolute value of the difference between the values of the opposing flows.
The direction of the net flow is determined by the sign of the difference.
If \( v_{\textrm{flow}}(i,j)>v_{\textrm{flow}}(j,i) \), the net flow value from \( i \) to \( j \) is given by  
\begin{equation}
v_\textrm{net}(i,j)=v_{\textrm{flow}}(i,j)-v_{\textrm{flow}}(j,i).
\end{equation}
For instance, if the flow from D to AV is larger than the flow in the opposite direction, then the net flow is from D to AV.

\subsection{General properties}
\begin{figure}
\includegraphics{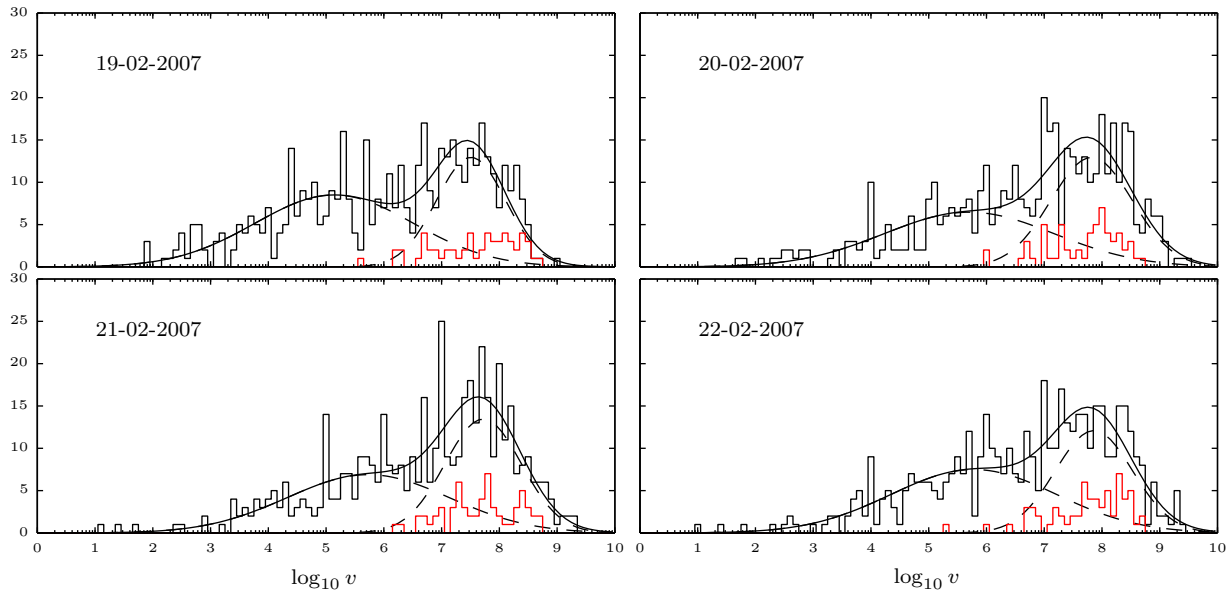}
\caption{The distribution of values of net nonloan flows (black histogram)
on a logarithmic scale with bin size \( \Delta\log_{10}v=0.1 \).
The components of the Gaussian mixture model are indicated with the dashed curves; the solid curve is the sum of the two components.
Net loan flows are overplotted in red.
The vertical axis counts the number of net flows per bin.}
\label{net value distribution}
\end{figure}

The distributions of net loan and nonloan flow values are presented in Figure~\ref{net value distribution}.
The parameters of the associated Gaussian mixture models are quoted in Table~\ref{GMMnet}.
The distribution of net nonloan flow values has the same general features as the distribution of the individual transactions.
However, unlike individual transactions, net flow values below {\au}\( 10^4 \) are rare;
net flows  around {\au}\( 10^8 \) are more prominent.

There are on average around 470 net nonloan flows each day.
Among these, roughly 110 consist of a single transaction and 50 consist of two transactions, mostly between small banks.
At the other extreme, net flows between the largest four banks (D, BP, AV, T) typically have more than \( 10^3 \) transactions per day each.
Overall, the distribution of the number of transactions per net flow is approximated well by a power law with exponent \( \alpha=-1.0\pm0.2 \):
\begin{equation}
N_\textrm{net}(n)\propto n^{\alpha},
\end{equation}
where \( N_\textrm{net}(n) \) is the number of net nonloan flows that consist of \( n \) transactions (\( n \) ranges from 1 to more than 1000).
This is consistent with the findings for Fedwire reported in \cite{bech2010} (see right panel of Fig.~14 in \cite{bech2010}).

There are roughly 60 net loan flows each day. 
As many as 40 consist of only one transaction.
On the other hand, a single net loan flow between two large banks may comprise more than 30 individual loans.
The distribution of the number of transactions per net loan flow is difficult to infer due to poor statistics,
but it is consistent with a power law with a steeper exponent, \( -1.4\pm0.2 \), than that of the nonloan distribution.
There are no net loan flows below {\au}\( 10^5 \) or above {\au}\( 10^9 \).
Comparing net loan and nonloan flows, it is obvious that net loan flows cannot compensate each and every net nonloan flow.
Not only are there fewer net loan flows than nonloan flows, 
but the total value of the former is much less than the total value of the latter.

\begin{table}
\centering
\begin{tabular}{ccccccc}
\hline
Date   & \multicolumn{3}{c}{Component 1} & \multicolumn{3}{c}{Component 2} \\
 & \( \mean{u} \) & \( \sigma_u^2 \) & \( P \) & \( \mean{u} \) & \( \sigma_u^2 \) & \( P \) \\
\hline
19-02-2007 & 5.14 & 1.88 & 0.60 & 7.51 & 0.36 & 0.40  \\
20-02-2007 & 5.70 & 2.17 & 0.51 & 7.82 & 0.50 & 0.49  \\
21-02-2007 & 5.73 & 1.97 & 0.52 & 7.72 & 0.44 & 0.48  \\
22-02-2007 & 5.78 & 2.06 & 0.57 & 7.86 & 0.45 & 0.43  \\
\hline
\end{tabular}
\caption{Mean \( \mean{u} \), variance \( \sigma_u^2 \), and mixing proportion \( P \) of the Gaussian mixture components
appearing in Figure~\ref{net value distribution}
(\( u=\log_{10}v \)).}
\label{GMMnet}
\end{table}

Net loan and net nonloan flows are not correlated; the correlation coefficient is 0.3.
Restricting net loan flows to those that have three transactions or more does not improve the correlation.
If a net loan flow  between two banks was triggered to a significant degree by the magnitude and the direction of
net nonloan flow between these bank, one expects a correlation between net loan and nonloan flows.
Our examination shows that in this respect loan flows are decoupled from nonloan flows.
The connection between them is indirect.
Namely, nonloan flows cause an imbalance in the account of each bank, which
is subsequently compensated by loan flows, which are largely unrelated to the nonloan flows that caused the imbalance.

\subsection{Degree distribution and assortativity}
\begin{figure}[t!]
{\centering
\subfloat[]{\includegraphics{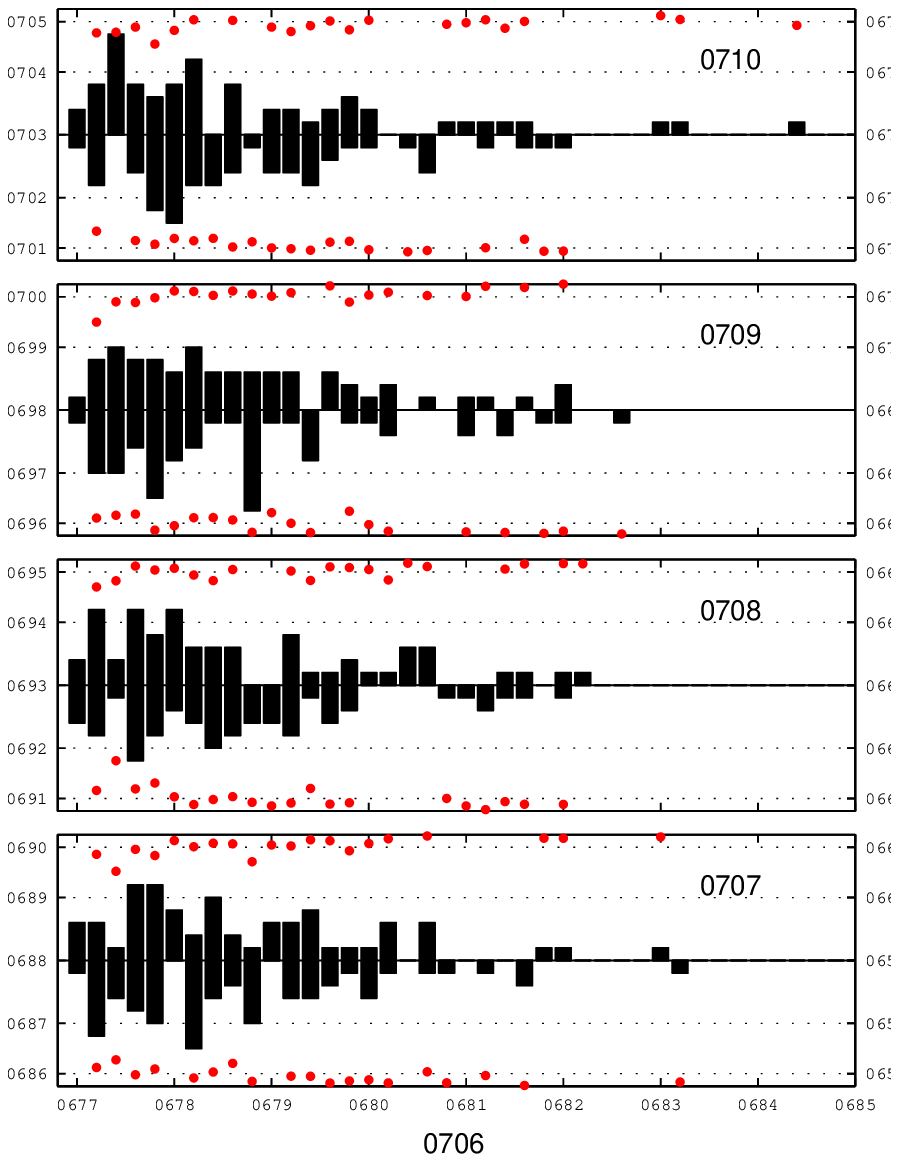}\label{degree distribution}}
\subfloat[]{\includegraphics{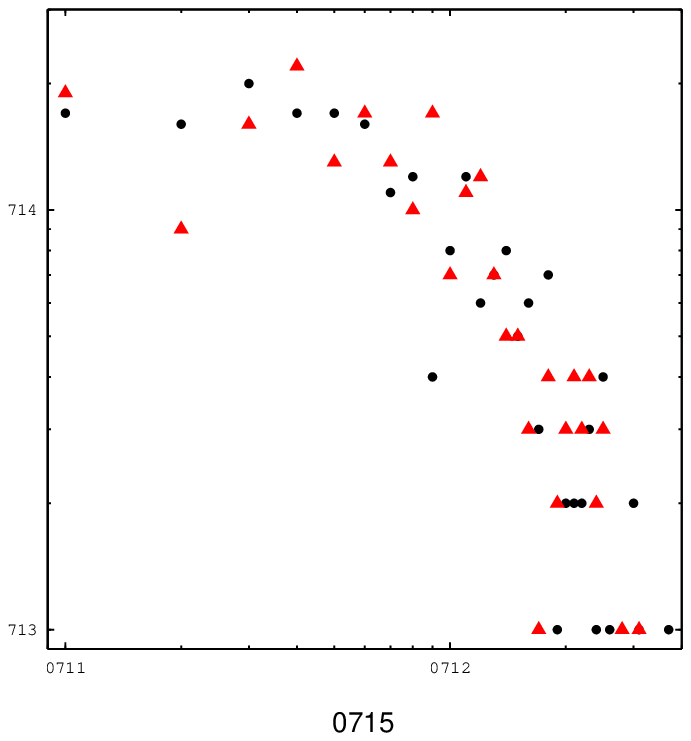}\label{combined degree distribution}}
}
\caption{\subref{degree distribution} 
Degree distribution of the net nonloan flow networks (for convenience, in-degrees are positive and out-degrees are negative).
The total value of the net flows corresponding to the specific degrees is shown with red dots
(the log of value in {\au}\( 10^9 \) is indicated on the right vertical axis).
\subref{combined degree distribution}
Degree distribution of the net nonloan flows
when the degree data for all four days are aggregated
(in-degrees are circles; out-degrees are triangles).
}
\end{figure}

We define the in-degree of node \( i \) as the number of net flows that terminate at \( i \), i.e.\ the number of net flows with destination \( i \), and
the out-degree as the number of net flows that originate from \( i \), i.e.\ the number of net flows with source \( i \).
The degree distribution of the nonloan networks is shown in Figure~\ref{degree distribution}.
Node BA has the highest in-degree of 37 on Monday, but on the other days it drops to 15 on average,
while the out-degree is 11.75 on average for this node.
The highest in-degrees are usually found among the four largest banks (D, BP, AV, T); the only exception is Monday, when AF's in-degree of 22 is greater than AV's 21,
and BA has the highest in-degree.
The highest out-degrees are usually achieved by D, BP, AV, T, W, and AH; the exceptions are Monday, when D's out-degree of 17 is less than AR's  and AP's 18,
and Thursday, when AV's out-degree of 16 is less than P's 18.

It is difficult to infer the shape of the degree distribution for individual days due to poor statistics.
The two-sample Kolmogorov-Smirnov (KS) test does not distinguish between the distributions on different days at the 5\% significance level.
With this in mind, we combine the in- and  out-degree data for all four days and graph the resulting distributions in Figure~\ref{combined degree distribution}.
We find that a power law distribution does not provides a good fit for either in- or out-degrees.
Visually, the distribution is closer to an exponential.
However, the exponential distribution is rejected by the Anderson-Darling test.

\begin{figure}[t!]
{\centering
\subfloat[]{\includegraphics{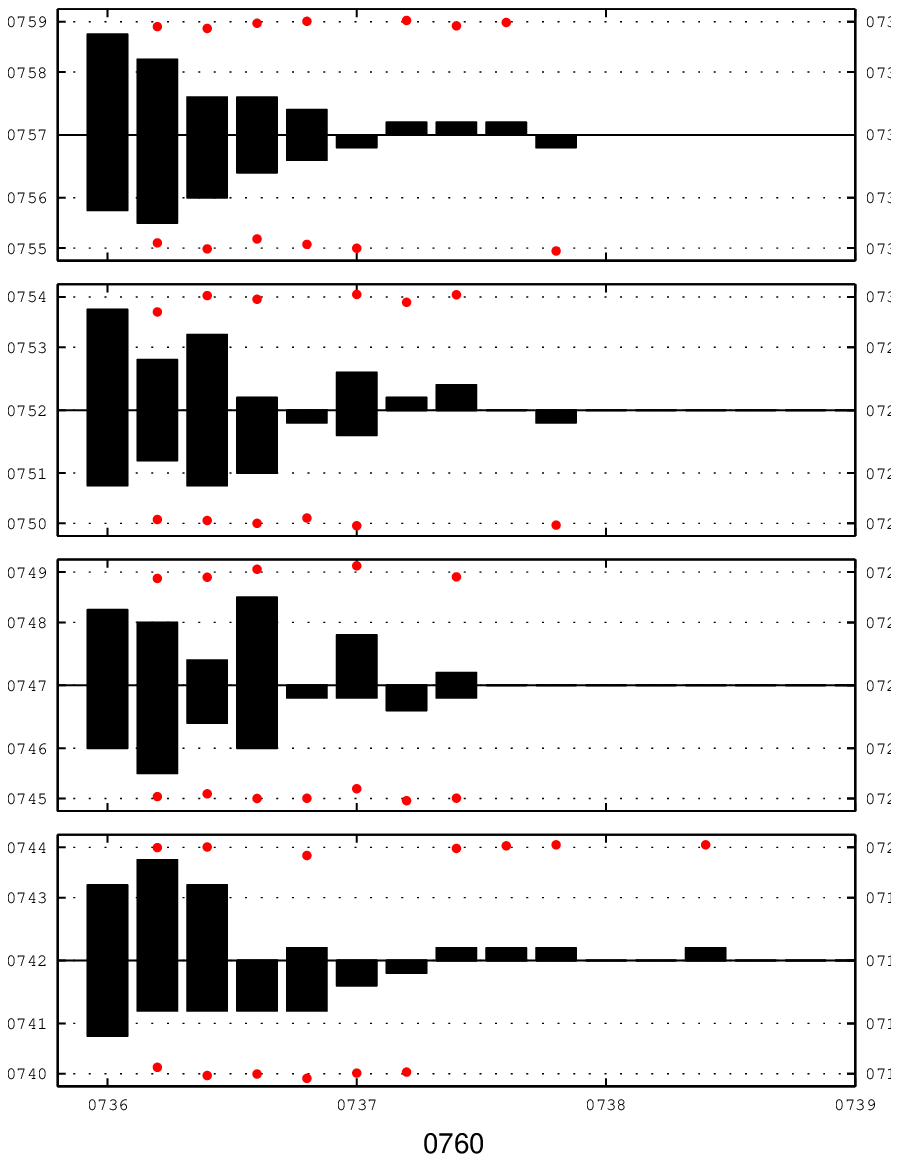}\label{degree distribution loans}}
\subfloat[]{\includegraphics{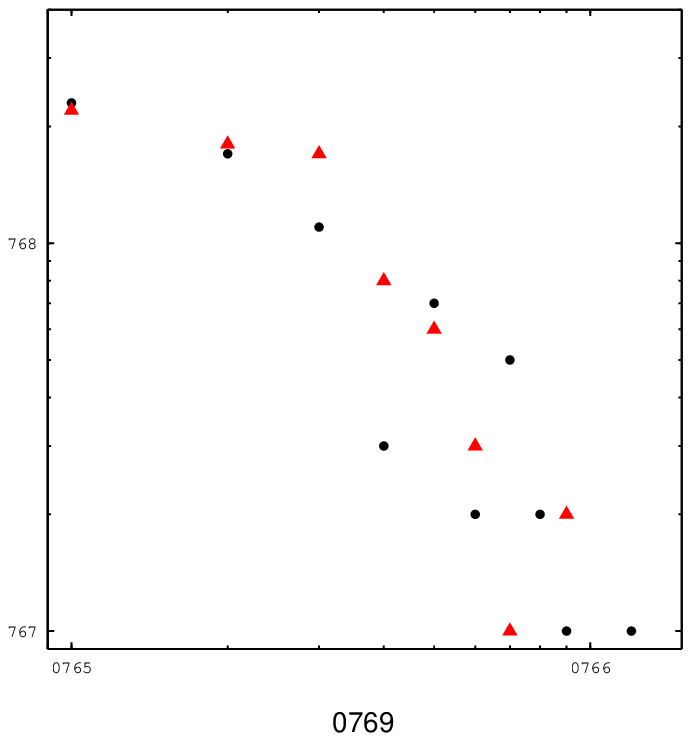}\label{combined degree distribution loans}}
}
\caption{\subref{degree distribution loans}
Same as Figure~\ref{degree distribution}, but for the net loan flow networks.
\subref{combined degree distribution loans}
Same as Figure~\ref{combined degree distribution}, but for the net loan flows.
}
\end{figure}

The degree distribution conceals the fact that flows originating or terminating in nodes of various degrees have different values
and therefore provide different contributions to the total value of the net flows.
Nodes with lower degrees are numerous, but the flows they sustain are typically smaller than
those carried by a few high-degree nodes.
In particular, for the nonloan flows, nodes with in-degree \( d\leq10 \) are numerous, ranging from 35 to 37, but their outgoing net flows carry about 20\% of the value on average.
On the other hand, nodes with \( d\geq17 \) are rare, but their flows carry 50\% of the value.
The same effect is observed for the out-degrees.

The degree distribution of the network of net loan flows is shown in Figure~\ref{degree distribution loans} 
(we ignore the nodes that have zero in- and out- degrees over four days). 
Similarly to nonloan flows, the KS test does not distinguish between the distributions on different days at the 5\% significance level.
The combined distribution is shown in Figure~\ref{combined degree distribution loans}.

To probe assortativity of the net flow networks, we compute the in-assortativity defined in \cite{piraveenan2010assortative} as
the Pearson correlation coefficient between the in-degrees of sources and destinations of the net flows (out-assortativity is computed similarly using the out-degrees).
The net nonloan flow network is disassortative, with in-assortativity of \( -0.39 \), \( -0.37 \), \( -0.38 \), \( -0.37 \)
and out-assortativity of \( -0.35 \), \( -0.38 \), \( -0.39 \), \( -0.37 \) on Monday, Tuesday, Wednesday, and Thursday, respectively.
The net loan flow network is less disassortative; the in-assortativity is \( -0.16 \), \( -0.26 \), \( -0.18 \), \( -0.19 \)
and the out-assortativity is \( -0.03 \), \( -0.10 \), \( 0.02 \), \( -0.20 \) for the same sequence of days.
In biological networks, the tendency of out-assortativity to be more assortative than in-assortativity  has been noted in \cite{piraveenan2010assortative}.

\subsection{Topology of the net flows}
\begin{figure*}[t!]
\includegraphics{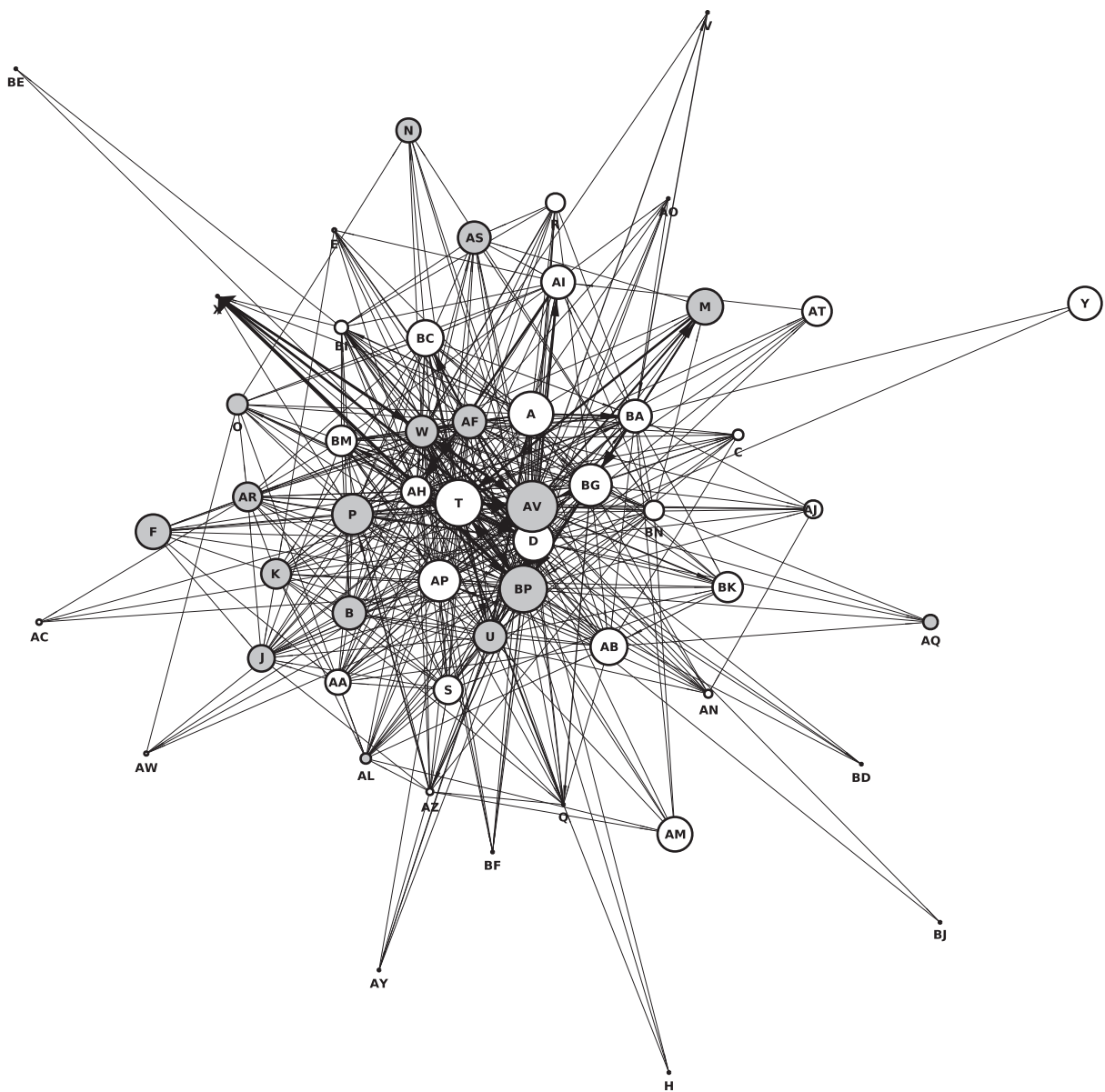}
\caption{Network of net nonloan flows on Tuesday, 20-02-2007.
White (grey) nodes represent negative (positive) imbalances.
The bank labels are indicated for each node.
The size of the nodes and the thickness of the edges are proportional to the logarithm of value of the imbalances and the net flows respectively.
}
\label{nonloan network 20}
\end{figure*}

Given the source and destination of each net flow, we can construct a network representation of the net flows.
An example of the network of net nonloan flows is shown in Figure~\ref{nonloan network 20}.
The size of the nodes and the thickness of the edges are proportional to the net imbalances
and net flow values respectively (on a logarithmic scale).
We use the Fruchterman-Reingold algorithm to position the nodes \cite{fruchterman1991graph};
the most connected nodes are placed in the centre, and the least connected nodes are moved to the periphery.
The core of the network is dominated by the four banks with the largest total value and the largest number of transactions: D, BP, AV, and T. 
The other big banks, such as AF, AH, and W, also sit near the core.
It is interesting to note the presence of several poorly connected nodes (Q, V, BF, and especially X) that participate 
in large incoming and outgoing flows, which produce only negligible imbalances in the banks themselves.

The sub-network consisting of D, BP, AV, BA, T, W, U, A, AH, AF, AP, and P
is fully connected on all five days, i.e.\ every node is connected to every other node.
The sub-network of D, AV, and BP is fully connected, even if we restrict the net flows to values above {\au}\( 10^8 \).

\begin{figure*}
{\centering
\subfloat[19-02-2007]{\includegraphics{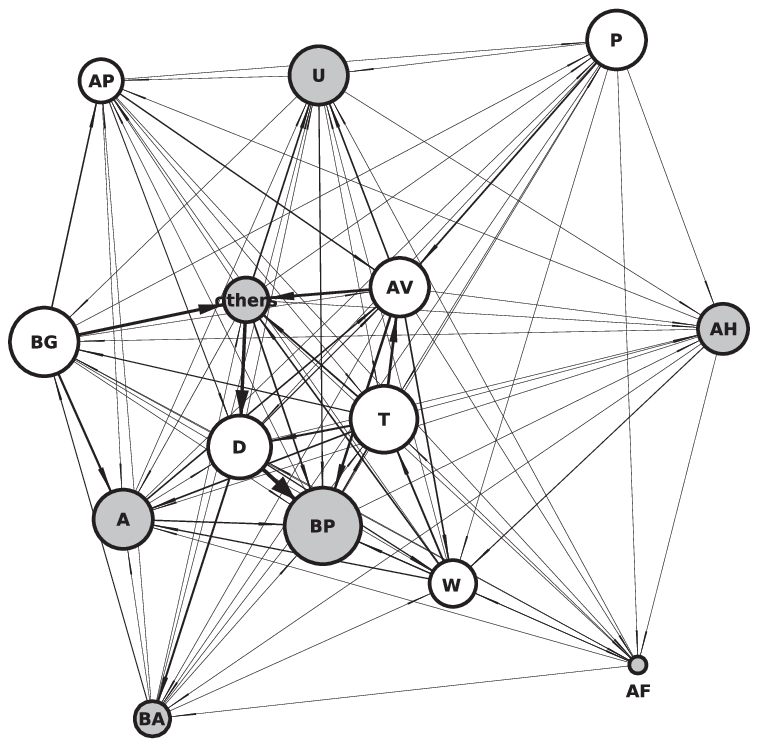} \label{restricted nonloan network 19}}
\makebox[0pt]{\raisebox{-2cm}[0cm][0cm]{\includegraphics{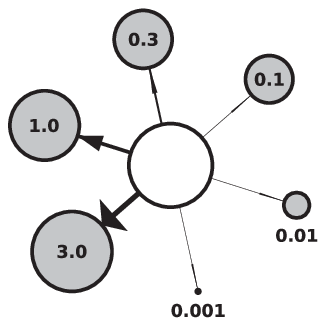}}}
\subfloat[20-02-2007]{\includegraphics{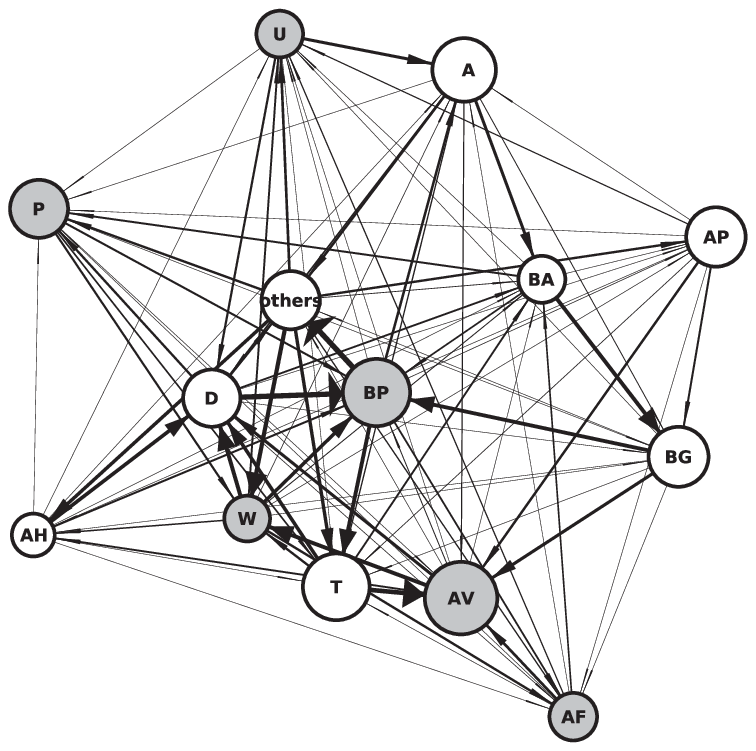} \label{restricted nonloan network 20}}\\
\subfloat[21-02-2007]{\includegraphics{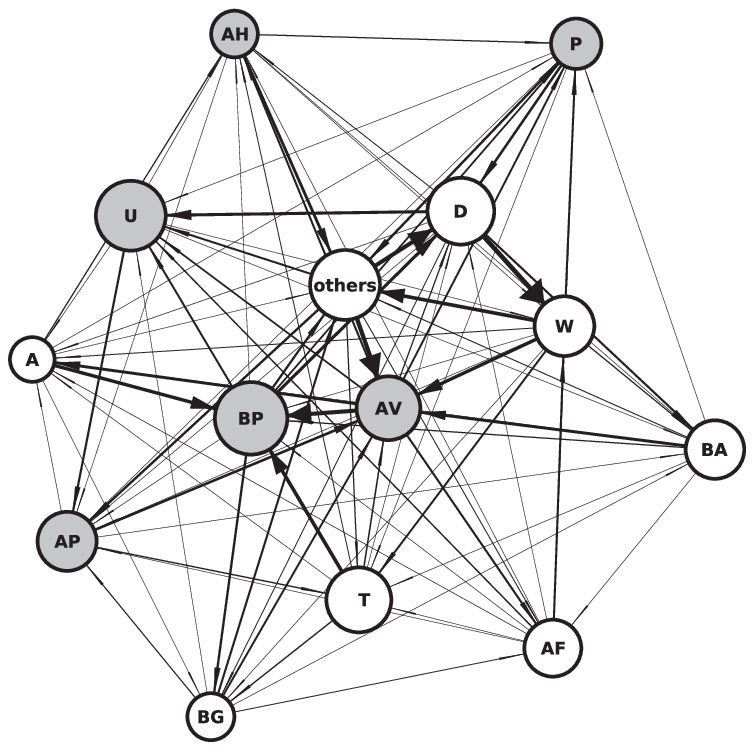} \label{restricted nonloan network 21}}
\subfloat[22-02-2007]{\includegraphics{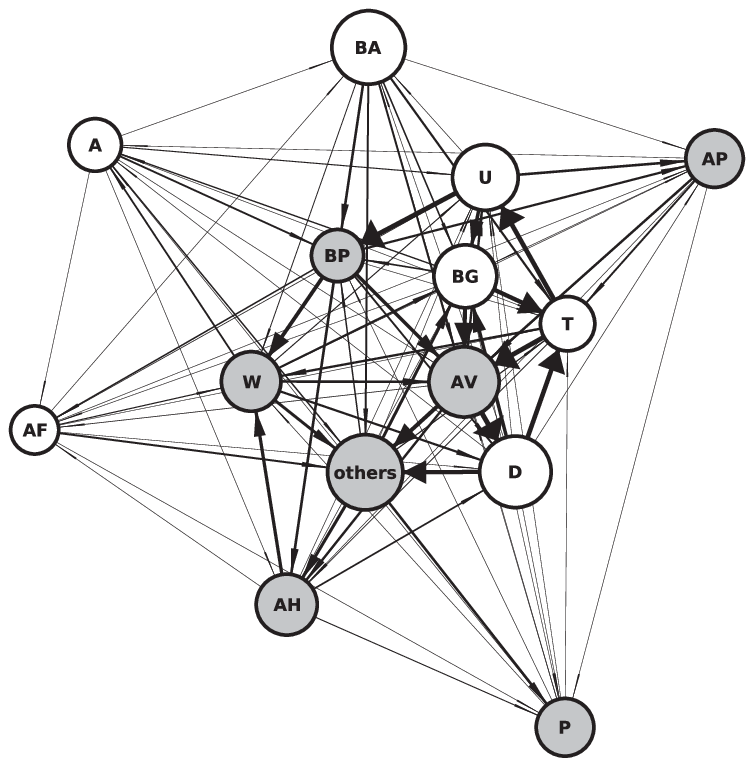} \label{restricted nonloan network 22}}
}
\caption{Networks of daily net nonloan flows for D, AV, BP, T, W, BA, AH, AF, U, AP, P, A, BG. 
All the other nodes and the flows to and from them are combined in a single new node called ``others''.
The size of the nodes and the thickness of the edges are proportional to the logarithm of value of the imbalances and the net flows respectively.
The value of the flows and the imbalances can be gauged by referencing a network shown in the middle, 
where the values of the flows are indicated in units of {\au}1 billion.
}
\end{figure*}

\begin{figure*}
{
\subfloat[19-02-2007]{\includegraphics{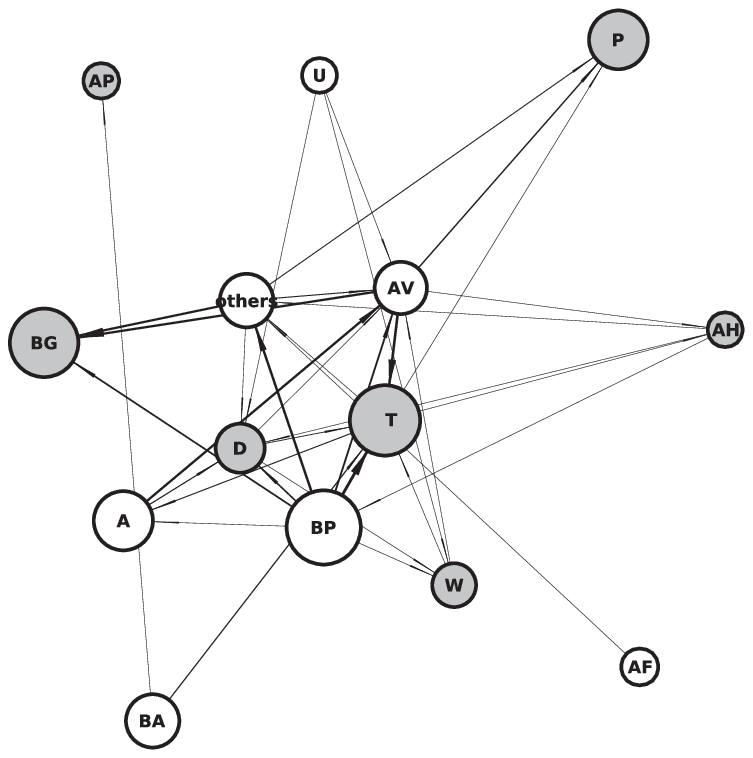} \label{loan network 19}}
\makebox[0pt]{\raisebox{-2cm}[0cm][0cm]{\includegraphics{test_flow_graph_arrows.eps}}}
\subfloat[20-02-2007]{\includegraphics{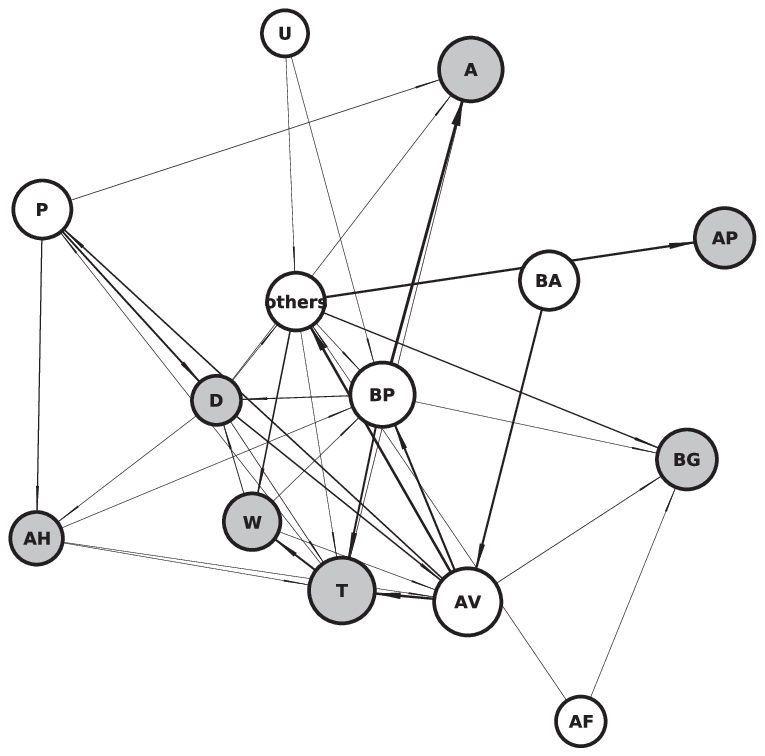} \label{loan network 20}}\\
\subfloat[21-02-2007]{\includegraphics{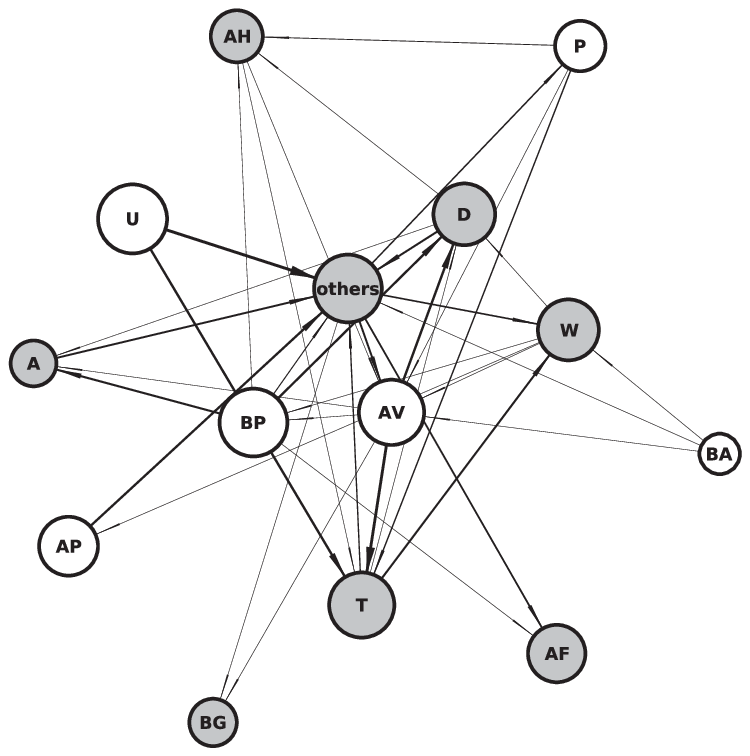} \label{loan network 21}}
\subfloat[22-02-2007]{\includegraphics{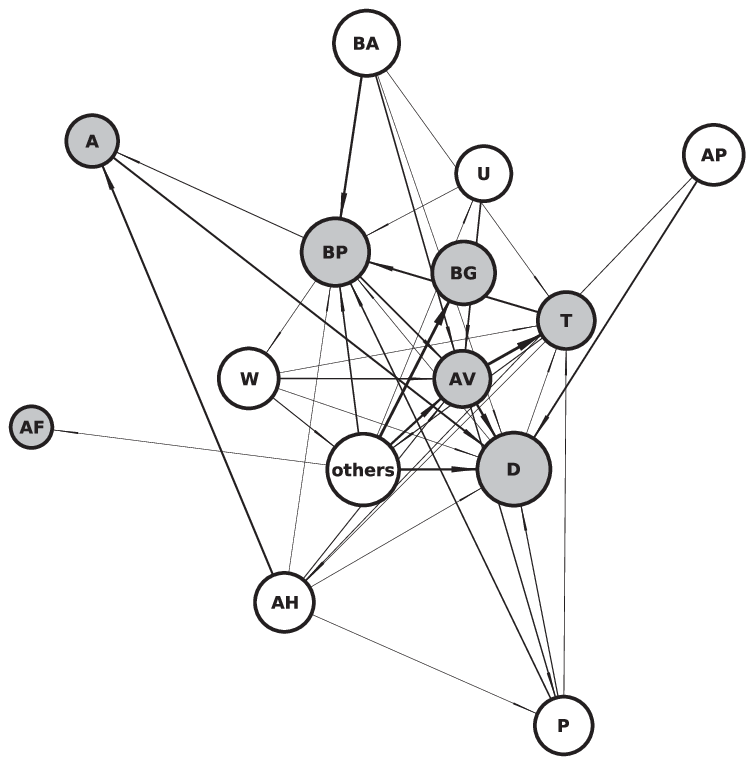} \label{loan network 22}}
}
\caption{Networks of daily net loan flows. The same nodes as in Figures~\ref{restricted nonloan network 19}--\ref{restricted nonloan network 22} are used.
The scale of the loan flows, the imbalances, and the positions of the nodes are the same as those used for the nonloan flows in Figures~\ref{restricted nonloan network 19}--\ref{restricted nonloan network 22}
 to simplify visual comparison.
}
\end{figure*}

In Figure~\ref{nonloan network 20}, the flows between the largest nodes are difficult to discern visually, because 
the nodes are placed too close to each other in the image.
We therefore employ the following procedure to simplify the network.
We consider the fully connected sub-network of twelve nodes, plus node BG, and combine all other nodes
into a new node called ``others'' in such a way that the net flows are preserved 
(BG is included because it usually participates in large flows and is connected to almost every node in the complete sub-network).
The result of this procedure 
applied to the daily nonloan networks is presented in Figures~\ref{restricted nonloan network 19}--\ref{restricted nonloan network 22}.
For these plots, we employ the weighted Fruchterman-Reingold algorithm, which  positions the nodes with large flows
between them close to each other.
The imbalances shown in Figure~\ref{restricted nonloan network 20} are the same as those of the full network in Figure~\ref{nonloan network 20}.
The daily networks of net loan flows for the same nodes are shown in Figures~\ref{loan network 19}--\ref{loan network 22}.

We observe that the largest flows on Monday (19-02-2007) were significantly lower than the flows on the subsequent days.
The largest nodes (D, BP, AV, T, W) are always placed close to the center of the network,
because they participate in the largest flows.
The topology of the flows is complex and difficult to disentangle, even if one concentrates on the largest flows (above {\au}\( 5\times10^8 \)).
For instance, on Monday, probably the simplest day, the flow of nonloans
is generally from BG to ``others'' to D to BP. 
There are also sizable flows from T to AV and from AV to ``others'' and BP.
However, lower value flows (below {\au}\( 5\times10^8 \)) cannot be neglected completely
because they are numerous and may contribute significantly to the imbalance of a given node.

Nodes D, T, BP, AV, and W form a complete sub-network of net loan flows on Monday, Tuesday, and Wednesday.
This sub-network is almost complete on Thursday too, except for the missing link between BP and W.
The appearance of the net loan network is different from that of the nonloan network, since
the same nodes participate in only a few loan flows.
Therefore, the position of a node in the network image is strongly influenced by the number of connections of that node.
Some of the poorly connected nodes are placed at the periphery despite the fact that they possess large flows.
The four largest nodes (D, T, BP, AV) are always positioned at the center of the network.

\subsection{Network variability}
The net nonloan flow network is extremely volatile in terms of flow value and direction.
For example, a {\au}\( 10^9 \) flow from D to BP on Monday transforms into
a {\au}\( 3.2\times10^9 \) flow in the same direction on Tuesday, only to be replaced by a {\au}\( 6.3\times10^8 \) flow in the opposite direction on Wednesday,
which diminishes further to {\au}\( 2.5\times10^9 \) on Thursday.
Nodes T and BP display a similar pattern of reversing flows between Tuesday and Wednesday. 
On the other hand, the net flow between T and AV maintains the same direction, but the flow value is strongly fluctuating.
In particular, a moderate {\au}\( 4.8\times10^8 \) flow on Monday rises to {\au}\( 1.9\times10^9 \) on Tuesday, then falls sharply to {\au}\( 2\times10^8 \) on Wednesday
and again rises to {\au}\( 2.2\times10^9 \) on Thursday.

Considering any three nodes, we observe that  circular and transitive flows are present on most days, 
the latter being more common.
The most obvious example is a circular flow between D, T, and BP on Thursday and
 a transitive flow involving BG, T, and AV on the same day.
The circular flows are unstable in the sense that they do not persist over two days or more.

The net loan flow network exhibits similar characteristics.
Few net loan flows  persist over the four days.
For example, the flow from AV to T has the same direction and is similar in value on all four days.
Circular loan flows are also present, as the flow between AV, T, and BP on Thursday demonstrates.

\section{Conclusions}

In this paper, we study the properties of the transactional flows between Australian banks participating in RITS.
The value distribution of transactions is approximated well by a mixture of two log-normal components, possibly reflecting
the different nature of transactions originating from SWIFT and Austraclear.
For the largest banks, the value distributions of incoming and outgoing transactions are similar.
On the other hand, the central bank displays a high asymmetry between the incoming and outgoing transactions,
with the former clearly dominating the latter for transactions below {\au}\( 10^6 \).

Using a matching algorithm for reversing transactions, we successfully separate transactions into loans and nonloans.
For overnight loans, we estimate the identification rate at 98\%.
The mean derived interest rate is within 0.01\% of the central banks' target rate of 6.25\%, while the standard deviation is about 0.07\%.
We find a strong anti-correlation between loan and nonloan imbalances (Pearson coefficient is about 0.9 on most days).
A likely explanation  is that nonloan flows create surpluses in some banks.
The banks lend the surplus to banks in deficit, creating loan flows that counteract the imbalances due to the nonloan flows.
Hence, loan and nonloan imbalances of individual banks are roughly equal in value and opposite in sign on any given day.

The flow networks are structurally variable, with 20\% of nonloan flows and 50\% of loan flows replaced every day.
Values of persistent flows, which maintain the same source and destination over at least two consecutive days, vary significantly from day to day. 
Some flow values change by several orders of magnitude.
Persistent flows increase in value several-fold between Monday and Tuesday.
Individual flow values can change by several orders of magnitude on the following day.
Overall, there is a reasonable correlation between the flow values on consecutive days (Pearson coefficient is 0.65 for nonloans and 0.76 for loans on average).
We also find that larger banks tend to sustain larger loan flows, in accord with the intuitive expectations.
However, there is no correlation between loan and nonloan flows.

We examine visually the topology of the net loan and nonloan flow networks.
The centre of both networks is dominated by the big four banks.
Twelve banks form a complete nonloan sub-network, in which each bank is connected to every other bank in the sub-network.
The three largest banks form a complete sub-network even if the net flows are restricted to values above {\au}\( 10^8 \).
Our examination reveals that the network topology of net flows is complicated,
with even the largest flows varying greatly  in value and direction on different days.

Our findings suggest a number of avenues for future research on interbank networks.
Firstly, the relationships we uncovered can be used to constrain analytical models and numerical simulations of interbank flows in financial networks.
In particular, our explanation of the link between the loan and nonloan imbalances needs to be tested in numerical simulations.
Secondly, it is necessary to analyse interbank markets in other countries to establish 
what elements of our results are signatures of general dynamics 
and what aspects are specific to the epoch and location of this study.
Even when high quality data are available, most previous studies concentrate on analysing static topological properties
of the networks or their slow change over time.
The internal dynamics of monetary flows in interbank networks has been largely ignored.
Importantly, one must ask whether the
strong anti-correlation between loan and nonloan imbalances is characteristic of RTGS systems whose institutional setup
resembles the Australian one or whether it is a general feature. 
For instance, in Italy a reserve requirement of 2\% must be observed on the 23rd of each month, which
may encourage strong deviations between loan and nonloan imbalances on the other days.

\section*{Acknowledgement}
We thank the Reserve Bank of Australia for supplying the data.
AS acknowledges generous financial support from the Portland House Foundation.

\section*{References}
\bibliographystyle{elsarticle-num}
\bibliography{asokolov.bib}

\begin{thebibliography}{10}
\expandafter\ifx\csname url\endcsname\relax
  \def\url#1{\texttt{#1}}\fi
\expandafter\ifx\csname urlprefix\endcsname\relax\def\urlprefix{URL }\fi
\expandafter\ifx\csname href\endcsname\relax
  \def\href#1#2{#2} \def\path#1{#1}\fi

\bibitem{kolaczyk2009statistical}
E.~Kolaczyk, Statistical analysis of network data: methods and models, Springer
  Verlag, 2009.

\bibitem{jackson2008social}
M.~Jackson, Social and economic networks, Princeton Univ Pr, 2008.

\bibitem{caldarelli2007scale}
G.~Caldarelli, Scale-Free Networks: Complex webs in nature and technology,
  Oxford University Press, USA, 2007.

\bibitem{schweitzer2009economic}
F.~Schweitzer, G.~Fagiolo, D.~Sornette, F.~Vega-Redondo, D.~White, Economic
  networks: What do we know and what do we need to know?, Advances in Complex
  Systems 12~(4-5) (2009) 407--422.

\bibitem{haldane2011}
A.~Haldane, R.~May, {Systemic risk in banking ecosystems}, Nature 469~(7330)
  (2011) 351--355.

\bibitem{may2008ecology}
R.~May, S.~Levin, G.~Sugihara, {Ecology for bankers}, Nature 451~(21) (2008)
  893--895.

\bibitem{newman2003mixing}
M.~Newman, {Mixing patterns in networks}, Physical Review E 67~(2) (2003)
  26126.

\bibitem{boss2004}
M.~Boss, H.~Elsinger, M.~Summer, S.~Thurner, {Network topology of the interbank
  market}, Quantitative Finance 4~(6) (2004) 677--684.

\bibitem{kyriakopoulos2009network}
F.~Kyriakopoulos, S.~Thurner, C.~Puhr, S.~Schmitz, Network and eigenvalue
  analysis of financial transaction networks, The European Physical Journal
  B-Condensed Matter and Complex Systems 71~(4) (2009) 523--531.

\bibitem{inaoka2004fractal}
H.~Inaoka, T.~Ninomiya, K.~Taniguchi, T.~Shimizu, H.~Takayasu, Fractal network
  derived from banking transaction--an analysis of network structures formed by
  financial institutions, Bank of Japan Working Papers 4.

\bibitem{de2006fitness}
G.~De~Masi, G.~Iori, G.~Caldarelli, Fitness model for the {Italian} interbank
  money market, Physical Review E 74~(6) (2006) 066112.

\bibitem{iori2007trading}
G.~Iori, R.~Ren{\'{o}}, G.~De~Masi, G.~Caldarelli, Trading strategies in the
  {Italian} interbank market, Physica A: Statistical Mechanics and its
  Applications 376 (2007) 467--479.

\bibitem{iori2008network}
G.~Iori, G.~De~Masi, O.~Precup, G.~Gabbi, G.~Caldarelli, A network analysis of
  the {Italian} overnight money market, Journal of Economic Dynamics and
  Control 32~(1) (2008) 259--278.

\bibitem{soramaki2007}
K.~Soram\"aki, M.~L. Bech, J.~Arnold, R.~J. Glass, W.~Beyeler, {The topology of
  interbank payment flows}, Physica A: Statistical Mechanics and its
  Applications 379~(1) (2007) 317--333.

\bibitem{bech2010}
M.~L. Bech, E.~Atalay, The topology of the federal funds market, Physica A:
  Statistical Mechanics and its Applications 389~(22) (2010) 5223--5246.

\bibitem{imakubo2010transaction}
K.~Imakubo, Y.~Soejima, The transaction network in {Japanese} interbank money
  markets, Monetary and Economic Studies 28 (2010) 107--150.

\bibitem{cajueiro2008role}
D.~Cajueiro, B.~Tabak, The role of banks in the {Brazilian} interbank market:
  Does bank type matter?, Physica A: Statistical Mechanics and its Applications
  387~(27) (2008) 6825--6836.

\bibitem{rordam2008topology}
K.~R{\o}rdam, M.~Bech, The topology of {Danish} interbank money flows, FRU
  Working Papers.

\bibitem{li2010}
S.~Li, J.~He, Y.~Zhuang, {A network model of the interbank market}, Physica A:
  Statistical Mechanics and its Applications.

\bibitem{gallagher2010}
P.~Gallagher, J.~Gauntlett, D.~Sunner, {Real-time Gross Settlement in
  Australia}, Reserve Bank of Australia Bulletin September (2010) 61.

\bibitem{mclachlan2000finite}
G.~McLachlan, D.~Peel, {Finite mixture models}, Wiley-Interscience, 2000.

\bibitem{furfine2003interbank}
C.~Furfine, {Interbank Exposures: Quantifying the Risk of Contagion.}, Journal
  of Money, Credit \& Banking 35~(1) (2003) 111--129.

\bibitem{ashcraft2007systemic}
A.~Ashcraft, D.~Duffie, Systemic illiquidity in the federal funds market, The
  American economic review 97~(2) (2007) 221--225.

\bibitem{rbarepos}
F.~Campbell, {Reserve Bank Domestic Operations under RTGS}, Reserve Bank of
  Australia Bulletin November (1998) 54.

\bibitem{piraveenan2010assortative}
M.~Piraveenan, M.~Prokopenko, A.~Zomaya, {Assortative Mixing in Directed
  Biological Networks}, IEEE IEEE/ACM Transactions on Computational Biology and
  Bioinformatics.

\bibitem{fruchterman1991graph}
T.~Fruchterman, E.~Reingold, Graph drawing by force-directed placement,
  Software - Practice and Experience 21~(11) (1991) 1129--1164.

\end{thebibliography}

\end{document}